\documentclass[a4paper,fleqn,usenatbib]{mnras}

\usepackage[T1]{fontenc}
\usepackage{ae,aecompl}
\usepackage{epstopdf}

\usepackage{graphicx}	
\usepackage{amsmath}	
\usepackage{amssymb}	
\usepackage{hyperref}
\usepackage{array}
\usepackage{dblfloatfix}
\usepackage[super]{nth}
\usepackage[utf8]{inputenc}
\usepackage{multirow}

\usepackage{ulem} 
\usepackage{color} 

\title[Statistics of collision parameters]{Statistics of collision parameters computed from 2D simulations}

\author[\'A. S\"uli]{
\'A. S\"uli,$^{1,2}$ \thanks{E-mail: a.suli@astro.elte.hu}
\\
$^{1}$Department of Astronomy, E\"otv\"os University, H-1117 Budapest, P\'azm\'any P\'eter s\'et\'any 1/A, Hungary\\
$^{2}$Konkoly Observatory, Research Centre for Astronomy and Earth Sciences, H-1121 Budapest, Konkoly Thege Mikl\'os \'ut 15-17, Hungary
}

\date{Accepted XXX. Received YYY; in original form ZZZ}

\pubyear{2015}

\begin{document}
\label{firstpage}
\pagerange{\pageref{firstpage}--\pageref{lastpage}}
\maketitle

\begin{abstract}
There are two popular ways to speed up simulations of planet 
formation via increasing the collision probability: ({\it i}) confine 
motion to 2D, ({\it ii}) artificially enhance the 
physical radii of the bodies by an expansion factor. In this paper I have 
performed 100 simulations each containing 
$10^4$ interacting bodies and computed the collision parameters
from the results of the runs. Each run was executed for a lower
and a higher accuracy parameter. The main goal is to determine 
the probability distribution functions of the collision parameters 
and their dependence on the expansion factor. A simple method is 
devised to improve the determination of the collision parameters 
from the simulation data. It was shown that the distribution of the 
impact parameter is uniform and independent of the expansion factor. 
For real collisions the impact velocity is greater than 
1 mutual escape velocity, a finding that can be explained using the 
two-body problem. The results casts some doubts on simulations of 
the terrestrial planets' final accretion that have assumed merge. 
Collision outcome maps were created adopting 
the fragmentation model of \cite{Leinhardt2012} to estimate the number 
of different types of collisions. A detailed comparison with earlier
works indicates that there are similarities as well as significant
differences between the different works. The results indicate that as 
the planetary disc matures and the masses of the bodies differs 
progressively than the majority of collisions lead to mass growth either 
via partial accretion or via graze-and-merge collision.
\end{abstract}




\section{Introduction}

According to the most widely accepted model, the nebular
hypothesis states that the formation process of the
terrestrial planets is the result of a series of three 
consecutive but partly overlapping stages 
\citep{Lissauer1993,Chambers2004,Morbidelli2012}. In the first
stage the micrometer-sized dust grains coagulates into small
aggregates through pairwise collisions \citep{Dullemond2005}.
The collisions may proceed further to form 1 - 100 km sized
bodies, which are called planetesimals. It must be noted that
this process making planetesimals can only be effective if the
conditions are suitable
\citep{Weidenschilling1977,Weidenschilling1997a}.
\cite{Weidenschilling1977} has described the dust motion
in protoplanetary discs and showed that dust grains from
micron to a few meters in size experience a radial
motion towards the star. This radial drift strongly depends on
the size and bodies with a critical size spiral into the star
in a fraction 
of the disc lifetime. This loss of material quickly clears the
disc preventing planetesimal formation. This depletion 
process is called the "radial-drift barrier" of planet
formation \citep{Laibe2012} and was first
studied in a minimum mass solar nebula in which the 
critical size corresponds to meter-sized bodies and thus later
was inaccurately referred to as the "meter-size barrier".

\par
There exist an alternative hypothesis, where planetesimals or
even Mars-size embryos may form directly from small dust
grains concentrated by turbulence followed by gravitational
collapse \citep{Johansen2009,Cuzzi2010}, or in pressure maxima
of protoplanetary disks \citep{Lyraetal2008}. Moreover, if
planetesimals are concentrated in a pressure maximum, they can
be quickly accreted by a growing embryo leading to the rapid
formation of a solid core and a giant planet
\citep{Lyraetal2008,Sandoretal2011,GuileraSandor2017}. In this
study I focus on the collision theory.

\par
Despite of the radial-drift barrier, the Solar System and
the observed exoplanets prove the existence of an efficient
and robust process that creates planetesimals in large
numbers. Once planetesimals have formed their gravitational
interactions controls further growth, which determines their
velocities and the parameters of occasional collisions. If
turbulence is not too vehement, than dynamical friction
ensures that the largest planetesimals have low relative
velocities. This leads to runaway growth in which the largest
objects grow more rapidly than smaller ones
\citep{Wetherill1989,Kokubo1996}.

\par
The next stage is the oligarchic phase when the largest
objects contain enough mass to dominate the velocity
dispersion of smaller planetesimals \citep{Ida1993}. In the
oligarchic growth mode the disc may partitioned into rings
where each ring is dominated by a single planetary embryo that
sweeps up planetesimals in its vicinity \citep{Kokubo1998}.
Nearby protoplanets grow at similar rates and their orbits
are separated by more than 5 Hill radius.

\par
Oligarchic growth ends when the number of planetesimal
drops below such a threshold where their damping effect on
protoplanet orbits become insufficient to prevent orbit 
crossing. This initiates the last phase of terrestrial
planet formation, involving the collision and mutual accretion
of protoplanets and embryos the so called giant impacts phase.
During this phase the number of embryos
declines and the final planets forge, in stable
non-crossing orbits. Most of the evolution took place after
the gas disc had dissipated, so the processes mainly consisted
of gravitational interactions and occasional collisions
between embryos.

\par
In the early stages of planet formation the vast number
of objects present in the dynamical system render the problem
unfeasible for direct $N$-body methods. But at the beginning
of the last stage the number of significant bodies drops to a
few $10^2$ to $10^3$ hence $N$-body integrations become
applicable for calculating the evolution of the system. These
methods are accurate and were used extensively to study the
final assembly of the terrestrial planets.

\par
According to previous studies
\citep{Chambers2001,Chambers2013,Liu2011} the timescale of
planetary systems formation is ${\sim}100$ Myr. Integrating the
motion of several $10^2$ to $10^3$ gravitationally interacting
bodies for ${\sim}100$ Myr still requires a huge computational
effort consequently two-dimensional (2D) model was frequently
used. Another method to speed up $N$-body simulations of planet
formation is to scale up the physical radii of the bodies by a 
factor. Both of these techniques increase the collision probability
between objects and thus reducing the computational time.

\par
The 2D integrations of planet formation \citep{Cox1980,Lecar1986,
Beauge1990,Alexander1998}
have notably contributed to our understanding of planetary accretion,
and motivated scholars to investigate the problem in more detail or
extend the model into 3D. On the other hand, there are findings
indicating that 2D simulations may not provide a feasible model
of a planet formation. According to \cite{Kokubo1996} runaway growth
takes place in 3D simulations but not in 2D, although this growth mode
might have played a key role in planet formation. Another result of 
\cite{Chambers1998} indicated that the time scales of collision in 2D
and 3D simulation differ significantly (a factor of ${\sim}10$) thus
the evolution of large bodies - the major concern in planet formation
 - follows different paths.

\par
The numerical $N$-body simulations that assumed perfect accretion
have been successful at reproducing the broad characteristics of the
terrestrial planets in our Solar System  
\citep{Wetherill1994,Chambers2001,Quintana2002,Raymond2004,
Raymond2006,Raymond2009,Obrien2006,Quintana2006,
Quintana2014,Quintana2016,Liu2011}.
These results were all based on a small (up to a dozen) number of 
realizations performed for each set of initial conditions. A larger set 
of 50 simulations of planet formation around the Sun was recently 
performed by \cite{Fischer2014}, who using the perfect-accretion model, 
demonstrated the need for a larger suite of simulations in order to 
infer results from a distribution of final planet configurations.

\par
To date the late stage has been mainly investigated by accurate $N$-body
simulations while collisions were modeled generally by perfect 
accretion. 
However, this oversimplified assumption breaks for collisions
with higher velocity and/or larger impact parameter \citep{Kokubo2010}.
This simplification was necessary in order to keep the number of bodies 
below the initial number and the lack of detailed models on 
collisions between planetary mass bodies. Having a better concept of
collisions and how it influences the evolution of the planets
is a primary key in understanding how planets emerge from a swarm
of planetesimals. For this very reason
\cite{Leinhardt2012}, hereafter LS12 have conducted high-resolution
simulations of collisions between planetesimals and the results were
used to isolate the effects of different impact parameters on collision 
outcome. Their model predicts the size and velocity distribution of 
fragments as a function of the impact velocity, impact angle and 
projectile to target mass ratio. 
The authors identified the boundaries between different types of 
collision: ({\it i}) simple merger, ({\it ii}) merger 
with some mass escaping as fragments, and ({\it iii}) hit-and-run 
collisions. The analytic model developed by LS12 is a powerful tool that 
has two major advantages: it improves the physics of collisions in 
numerical simulations of planet formation and collision evolution, and 
is easy to adapt to an $N$-body code.

\par
The collision model of LS12 made
possible to implement more sophisticated, inter-particle gravity enabled
$N$-body simulations of late-stage planet formation. 
\cite{Lines2014} used
this model to simulate planet formation in the Kepler-34(AB) system's
circumbinary protoplanetary disk to examine whether planets can form in
a hostile environment. The same collision model for gravity-dominated 
bodies has been implemented into an $N$-body tree code 
\citep{Bonsor2015} and used to examine planet formation.

\par
\cite{Chambers2013} implemented this comprehensive collision model
into the widely used ${Mercury}$ integration package 
\citep{Chambers2001}. 
Eight simulations of planet formation using the new collision model 
were presented and compared to eight simulations that were previously 
performed using the perfect-accretion collision model 
\citep{Chambers2001}. 
The new simulations form 3 to 5 terrestrial planets moving on widely 
spaced orbits with growth complete by 400 My. The final planets that 
formed in each of these sets were shown to be comparable despite the 
significant difference in the number and frequency of collisions. 
In addition, the accretion timescales were about twice as long when 
fragmentation was included.

\par
Terrestrial planet formation are characterized by countless collisions 
among bodies highlighting that collisions are the core agent of planet 
formation. Impacts outcome span multiple regimes depending on the 
collision parameters. The collision history of the final body can 
largely influence the planet's growth, stability, bulk composition, 
and habitability \citep{Chambers2001,Chambers2013,Bonsor2015}.

\par
In this paper I present the results of 100 simulations of 
planet formation in 2D started from the beginning of the 
stochastic stage. The main focus of the work is to present a detailed 
and comprehensive picture of the statistical distribution 
of the collision parameters. The dependence of these statistics 
on the expansion factor is also studied and a comparison of the results 
with others is presented. This is a first step towards examining 
the statistical distribution of the collision parameters. 
With a reliable statistics one can assess the frequency of
different collision outcome and posterior estimate the
number of collisions lead to perfect merging. This is 
essential since in the majority of the $N$-body simulations 
colliding bodies were assumed to merge into a single new object,
however this basic premise has not been thoroughly and 
deliberately tested. This is a very strong assumption 
and can largely bias the result of planet formation, thus
there is an impetus to assess the occurrence of perfect merging. 
The presented statistics can be also useful for $N$-body modelers
to preform Monte Carlo simulations of planet formation including
the model of LS12 \citep{Stewart2012}.

\par
Using the scaling laws derived by LS12 \cite{Stewart2012}
presented a retrospective analysis of previous simulations 
to estimate the range of true collision outcomes. In this
work a similar analysis is performed on the collision data
adopting the LS12 model to estimate the occurance of collision
outcomes and the results are compared to previous works.

\par
The rest of this paper is organized as follows. Section 2 contains
a description of the simulations and initial conditions. Section 3 
describes the results of these simulations, including the 
speed of the runs, the description of the collision geometry, 
the collision data and the distribution of the collision
parameter. Section 4 contains a simple analytic model which is used
to discuss the observed impact velocity distribution. Section 5 shows
the collision outcome maps derived by the model of LS12 and the 
collisions of the simulations are plotted on the maps. A comparison 
with other works are also presented and the main results are summarized 
in Section 6.

\section{The simulations}
\label{section_simulations}

The equations of motion of $10^4$ 
protoplanets around a star in the barycentric coordinate system $Oxy$ was
integrated. The protoplanets are confined to 2D and placed initially in the 
terrestrial region extending from 0.5 to 1.5 au. In this model each body 
was interacting with all the other bodies and all bodies were modeled 
as a sphere.

\par
In order to determine the total mass of solids in the ring the 
parameters of the minimum mass solar nebula \citep{Hayashi1981} was used 
in which the surface density of solids is:
\begin{equation}
\Sigma_{\mathrm{s}} = \Sigma_1 r^{-\frac{3}{2}},
\label{Eq_surfacedensity}
\end{equation}
where $\Sigma_1 = 7\, \mathrm{g cm^{-2}}$ is the surface density of 
solids at 1 au and $r$ is the distance from the star. 
Integrating Eq. (\ref{Eq_surfacedensity}) from 0.5 to 1.5 au 
results in 
$m_{\mathrm{tot}} = 5.123 \times 10^{-6} M_\odot = 1.71 M_\oplus$ 
where $M_\odot$ and $M_\oplus$ denote the mass of the Sun and the Earth, 
respectively.

\par
The simulations were started with $N_{\mathrm{p}} = 10^4$ bodies around 
a star with 1 $M_\odot$. Each protoplanet has the same physical properties: 
$m_{\mathrm{p}} = m_{\mathrm{tot}}/N_{\mathrm{p}} = 5.123 \times 10^{-10} M_\odot
\approx M_\mathrm{Ceres}$ and the density is 
$\rho_{\mathrm{p}} = 2.0$ $\mathrm{g cm^{-3}}$ which corresponds 
to the mean density of Ceres and consistent with silicon-rich 
rocky bodies. Consequently the radius of the bodies $R_{\mathrm{p}}$ 
is approximately 500 km (see Table \ref{Tab_01}).

\par
The semi-major axes $a$ for this number of protoplanets are generated 
randomly with probabilities weighted in order 
to reproduce the disc surface density profile as defined by 
Eq. (\ref{Eq_surfacedensity}). Eccentricities $e$ are randomized from 
a Rayleigh distribution with rms values of 0.02 with an upper limit 
of 0.2. The mean anomaly $M$ and argument of pericenter $\omega$ 
are randomized uniformly from $[0, 2\pi]$.

\begin{table}
\centering
\caption{The mass, radius and the density of the protoplanets. In the 
\nth{2} column the values are given in au and solar mass. In 
the \nth{3} column the values are given in Earth mass and radius units, 
while in the last column in SI units.}
\begin{tabular}{llll}
\hline
\hline
$m_{\mathrm{p}}$    & 5.1230212063 $\times 10^{-10}$ & 1.71  $\times 10^{-4}$ & 1.01903 $\times 10^{21}$ \\
$R_{\mathrm{p}}$    & 3.3120486271 $\times 10^{-6}$  & 7.77 $\times 10^{-2}$ & 4.95475 $\times 10^{5}$ \\
$\rho_{\mathrm{p}}$ & 3.3662582520 $\times 10^{6}$   & - & 2.0 $\times 10^{3}$ \\
\hline
\end{tabular}
\label{Tab_01}
\end{table}

\par
In the simulations the gas disk had dissipated by this stage and
all collisions were assumed to be perfectly 
inelastic, forming a new body by conserving mass and linear momentum. 
Most of the previous work on planet formation used this oversimplified 
treatment of collisions.

\par
The $N$-body systems are stochastic and so a large number of simulations 
of a given system, with changes in the initial conditions, are required 
in order to reach relevant conclusions. Most numerical simulations designed 
to explore the stochastic stage lack the large number of 
realizations needed to account for the stochastic nature of $N$-body 
systems. I improve on this limitation by performing 100 simulations of 
planet formation around a Sun-like star. To study the dependence on the
expansion factor $f$ used to artificially enhance the physical radii 
of the bodies I have preformed 5 different set of simulation 
for $f = 1,\,2,\,3,\,5,\,10$. In order to have statistically meaningful 
results 10 sets of initial conditions were generated, denoted by 
run id = $1,\,2,\ldots,10$. For each value of $f$ all these 10 
initial condition were integrated for $T = 10^6$ years. The integrator 
used an adaptive step size with a tolerance of $\epsilon = 10^{-10}$ 
and $\epsilon = 10^{-13}$ for all 50 cases, resulting in total 100
simulations.

\par
The applied integrator is the Runge-Kutta-Fehlberg 7(8) algorithm with 
adaptive step size, which has an acceptable speed and accurate in 
most situations \citep{Fehlberg1968}. All simulations were performed 
with an open source GPU code 
{\tt red.cuda}\footnote{\url{https://github.com/suliaron/red.cuda}}
designed to integrate 
planet and planetesimal dynamics in the framework of the core accretion 
planet formation. The {\tt red.cuda} is written in CUDA C and 
runs on all NVIDIA GPUs with compute capability of at least 2.0. All simulation 
were performed on Tesla K20Xm device containing 2688 CUDA Cores and uses 
CUDA driver version and run-time version 7.0 and 6.5, respectively.

\section{Results}
\label{section_results}

\subsection{Speed of simulation run}
\label{section_speed_of_sim}

\begin{figure}
\includegraphics[width=0.95\columnwidth,keepaspectratio=true]{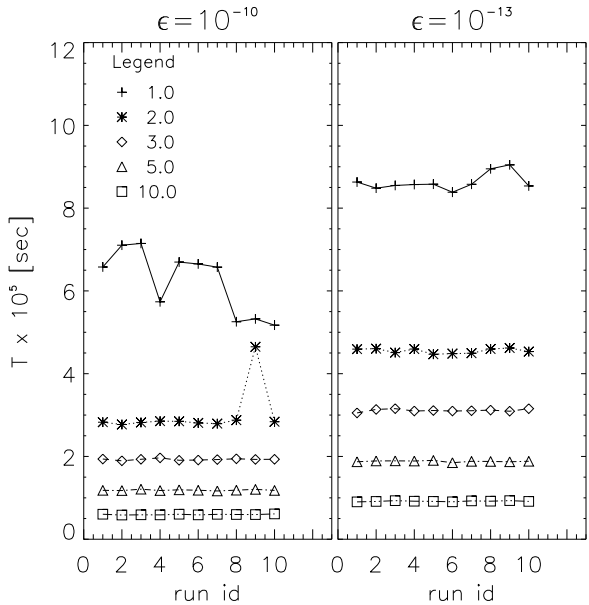}
\caption{The run-time of all the $2\times 50$ simulations for the 
two accuracy parameters. The data are plotted for $f=1, 2, 3, 5, 10$ 
as plus sign, asterisk, diamond, triangle, and square symbols, respectively.}
\label{Fig_01}
\end{figure}

I show how the typical CPU run-time of a simulation denoted by $T$ 
depends on $f$. Fig. \ref{Fig_01} displays $T$ for 
each $f$ value and for each of the 10 different runs. In the 
left panel the data correspond for the $\epsilon = 10^{-10}$ 
accuracy level, while the right one for $\epsilon = 10^{-13}$. The 
statistics of the data are given in Table \ref{Tab_02} where the 
minimum, maximum, mean and standard deviation of $T$ is calculated 
from the sample of the 10 runs belonging to a given $f$. The standard 
deviations, denoted by sd are usually more useful to describe the 
variability of the data. From Fig. \ref{Fig_01} and Table \ref{Tab_02}
it is clear that the variation of the run-time is the most significant 
for $f = 1$ and it gradually diminishes as $f$ gets larger.

\par
\begin{figure}
\includegraphics[width=0.95\columnwidth,keepaspectratio=true]{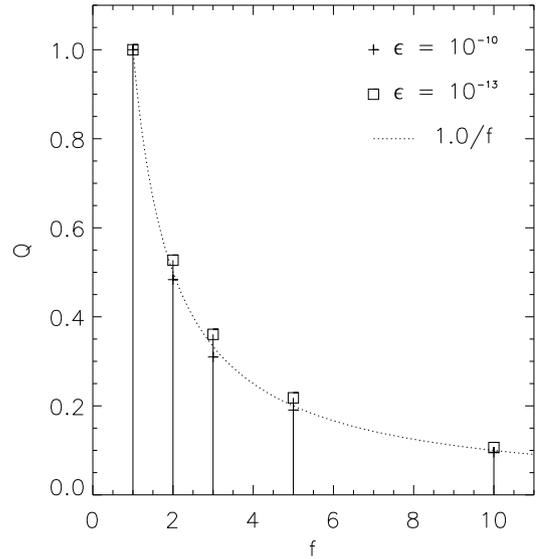}
\caption{The variation of $Q$ as a function of $f$ for $\epsilon = 
10^{-10}$ is plotted with plus sign and for $\epsilon = 10^{-13}$ 
plotted with square. The curve of $1/f$ is plotted by dotted line.}
\label{Fig_02}
\end{figure}

The other observable fact is the decrease of $T$ as $f$ increases. 
Each mean run-time $\overline{T}$ (\nth{4} and \nth{8} columns in Table 
\ref{Tab_02}) were divided by the maximum of the means:
\begin{equation}
Q(f) = \frac{\overline{T}(f)}{\mathrm{max}(\overline{T}(f))} = 
\frac{\overline{T}(f)}{\overline{T}(1)},
\label{Eq_Q}
\end{equation}
and the ratios $Q$ were plotted in Fig. \ref{Fig_02} as a function of 
$f$ for both accuracy values. From the figure it is clear that $Q$
is approximately inversely proportional to $f$, and the $1/f$
function (dotted curve) is apparently a very good 
approximation of the measured data. In 2D one can state that increasing 
the collision cross section of the bodies by $f$ will reduce the 
simulation time by $f$.

\par
The $q$ value in Table \ref{Tab_02} is calculated for each $f$ as
\begin{equation}
q(f) = \frac{\overline{T}(f; \epsilon = 10^{-13})}{\overline{T}(f; \epsilon = 10^{-10})},
\label{Eq_q}
\end{equation}
which has a mean of 1.527, indicating a 53\% increase in the mean 
run-time for a 3 orders of magnitude higher accuracy. 

\begin{table*}
\centering
\caption{The run-time in seconds for $
\epsilon = 10^{-10}$ and $\epsilon = 10^{-13}$. In the 
\nth{2} column the minimum, in the \nth{3} the maximum, in the \nth{4} 
the mean and in the \nth{5} the standard deviation of the $T$ is 
given for $\epsilon = 10^{-10}$. The same values are given in \nth{6} - 
\nth{9} for $\epsilon = 10^{-13}$. In the last column $q$ is the ratio 
between the means, see Eq. (\ref{Eq_q}).}
\begin{tabular}{rrrrr|rrrrr}
\hline
 & \multicolumn{4}{c}{$\epsilon = 10^{-10}$} & \multicolumn{4}{c}{$\epsilon = 10^{-13}$} & \\
$f$ & min($T$) & max($T$) & $\overline{T}$ & sd($T$) & min($T$) & max($T$) & $\overline{T}$ & sd($T$) & $q$ \\
\hline
\hline
    1 & 5.172$\times 10^{5}$ & 7.147$\times 10^{5}$ & 6.224$\times 10^{5}$ & 7.727$\times 10^{4}$ & 8.384$\times 10^{5}$ & 9.046$\times 10^{5}$ & 8.630$\times 10^{5}$ & 2.061$\times 10^{4}$ & 1.39 \\
    2 & 2.772$\times 10^{5}$ & 4.650$\times 10^{5}$ & 3.009$\times 10^{5}$ & 5.773$\times 10^{4}$ & 4.469$\times 10^{5}$ & 4.620$\times 10^{5}$ & 4.550$\times 10^{5}$ & 5.753$\times 10^{3}$ & 1.51 \\
    3 & 1.895$\times 10^{5}$ & 1.966$\times 10^{5}$ & 1.928$\times 10^{5}$ & 1.990$\times 10^{3}$ & 3.051$\times 10^{5}$ & 3.155$\times 10^{5}$ & 3.112$\times 10^{5}$ & 3.118$\times 10^{3}$ & 1.61 \\
    5 & 1.167$\times 10^{5}$ & 1.209$\times 10^{5}$ & 1.186$\times 10^{5}$ & 1.330$\times 10^{3}$ & 1.849$\times 10^{5}$ & 1.905$\times 10^{5}$ & 1.882$\times 10^{5}$ & 1.536$\times 10^{3}$ & 1.59 \\
   10 & 5.843$\times 10^{4}$ & 6.117$\times 10^{4}$ & 5.965$\times 10^{4}$ & 8.770$\times 10^{2}$ & 9.024$\times 10^{4}$ & 9.330$\times 10^{4}$ & 9.177$\times 10^{4}$ & 1.040$\times 10^{3}$ & 1.54 \\
\hline
\end{tabular}
\label{Tab_02}
\end{table*}

\begin{figure}
\includegraphics[width=0.95\columnwidth,keepaspectratio=true]{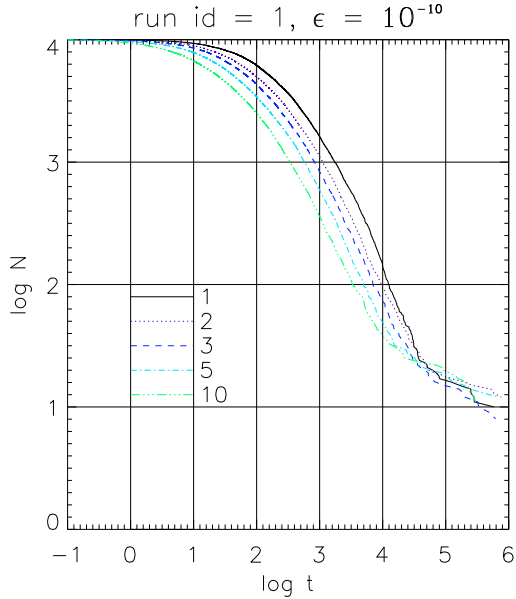}
\caption{The logarithm of the number of protoplanets as a function of 
the logarithm of time. The $f=1$ case is shown by black solid line, 
$f=2$ dotted purple, $f=3$ blue dashed, $f=5$ light blue dot-dashed 
and the $f=10$ green 3 dotted-dashed line.}
\label{Fig_03}
\end{figure}

The run time can be written as
\begin{equation}
T(f) = \sum_{i=0}^{n}T_i(f) = \eta \sum_{i=0}^{n}N^2(t_i,f)\delta t_i,
\label{Eq_runtime}
\end{equation}
where $n$ is the number of collisions, $N(t_i,f)$ is the number of 
bodies at time $t_i$, $\delta t_i$ is the simulation time between the $i
$th and $(i+1)$th collision and $\eta$ is a parameter characterizing the 
hardware. The number of bodies versus time is depicted in Fig. 
\ref{Fig_03} for different $f$ on logarithmic scale. We have an
almost identical figure for $\epsilon = 10^{-13}$. These curves are 
piecewise constant functions since each collision reduce 
the number by 1. It is visible that the curves in the $t\le 1$ and 
$t\ge 10^5$ yr intervals are similar and in between their behavior are 
alike, but the bigger $f$ is the smaller the time when the fast decrease 
begins. These curves match very well with previous 2D simulations e.g. 
\cite{Lecar1986,Alexander1998}. According to Eq. (\ref{Eq_runtime})
the $t\le 1$ part hardly contributes to the total run time since it is 
only 1 yr ($\delta t_i$ is small), while in the $t \ge 10^5$ interval 
$N$ is a few dozens for all $f$ values therefore it takes the same amount 
of time to complete.

\par
\begin{figure}
\includegraphics[width=0.95\columnwidth,keepaspectratio=true]{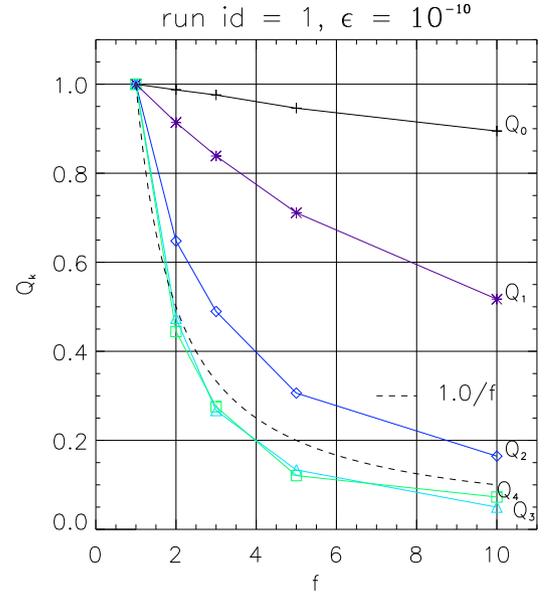}
\caption{The variation of $Q_k$ as a function of $f$. The curve of $1/f$ is plotted by dashed line.}
\label{Fig_04}
\end{figure}

In order to demonstrate the dependence of the run time on $f$ shown 
in Fig. \ref{Fig_02} let's write the ratio of the instantaneous run time 
belonging to $f=1$ and $f$ for specific simulation time epochs as 
follows 
\begin{equation}
Q_k(f) = \frac{T(\tau_k, f)}{T(\tau_k, 1)} \approx 
\left( \frac{N(\tau_k, f)}{N(\tau_k, 1)}\right)^2, \quad k=0,\ldots,4,
\label{Eq_Qk}
\end{equation}
where $\tau_k=10^k$. The results are depicted in Fig. \ref{Fig_04} for
the five different time epochs. We have an almost identical figure 
for $\epsilon = 10^{-13}$. It is clearly visible that with 
increasing $t$ the $Q_k$ ratios fit better and better to the $1/f$ 
function. The best fit is for $k=2,3$ and 4 (the $Q_3$ and $Q_4$
curves overlap). According to Eq. (\ref{Eq_runtime}) and Fig. 
\ref{Fig_03} it is obvious that in these cases both $N$ and 
$\delta t$ are large, therefore principally this interval determines 
the run-time. This reasoning is not an exact proof why the run time 
depends on the reciprocal of $f$ but clearly demonstrates the underlying 
cause, i.e. the specific decrease of $N$ as shown in Fig. \ref{Fig_03}.

\subsection{Collision geometry and parameters}
\label{section_coll_geom}

\begin{figure}
\includegraphics[width=0.95\columnwidth,keepaspectratio=true]{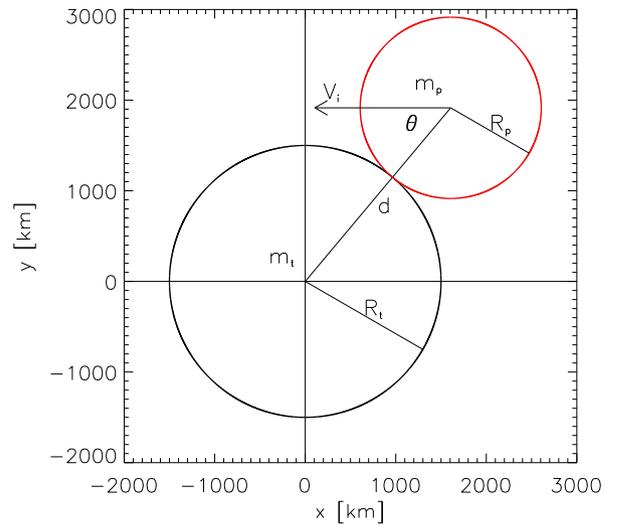}
\caption{Schematic of the collision geometry.}
\label{Fig_05}
\end{figure}

In the model initially all bodies have the same mass, but as collisions take 
place the masses will differ. Fig. \ref{Fig_05} shows the geometry of a 
collision. In what follows the target is the body with the larger mass 
$m_{\mathrm t}$ and the other body is called the projectile with mass 
$m_{\mathrm p}$. In the figure the target is stationary and the 
projectile is moving from right to left with speed $V_{\mathrm i}$, 
the units on the $x$ and $y$ axes are in km. The impact angle $\theta$, 
is defined at the time of first contact as the angle between the line 
connecting the centers of the two bodies and the impact velocity vector 
(when $\theta = 0^{\circ}$ the collision is refereed as head-on, while 
the $\theta = 90^{\circ}$ case corresponds to a grazing collision). 
The impact parameter $b$ is 
\begin{equation}
b = \sin\theta = B/d,
\label{Eq_impact_param}
\end{equation}
where $B$ is the $y$ coordinate of the projectile's centre and $d$ is 
distance between the centers of the bodies.

\par
The impact parameter has a notable influence on the collision outcome, 
because the kinetic energy of the projectile may only partially 
intersect the target for oblique impact, e.g. in the collision 
geometry shown in Fig. \ref{Fig_05}, the top of the projectile 
does not directly hit the target. 
Consequently a portion of the projectile may shear off and only 
the kinetic energy of the lower part  
of the projectile will be involved in disrupting the target. According 
to LS12 the outcome of a collision is dependent on the kinetic energy 
of the interacting mass which strongly depends on $b$.

\par
In order to describe the dependence of catastrophic disruption on 
impact angle, two geometrical collision groups were introduced by 
LS12: non-grazing, where most of the projectile interacts with the 
target and grazing where less than half the projectile interacts 
with the target. Following \cite{Asphaug2010}, the critical impact 
parameter
\begin{equation}
b_{\mathrm{crit}} = \frac{R_{\mathrm{t}}}{R_{\mathrm{t}} + R_{\mathrm{p}}},
\label{Eq_b_crit}
\end{equation}
is reached when the centre of the projectile  
is tangent to the surface of the target. 
Grazing impacts are defined to occur when $b > b_{\mathrm{crit}}$.

\begin{figure}
\includegraphics[width=0.95\columnwidth,keepaspectratio=true]{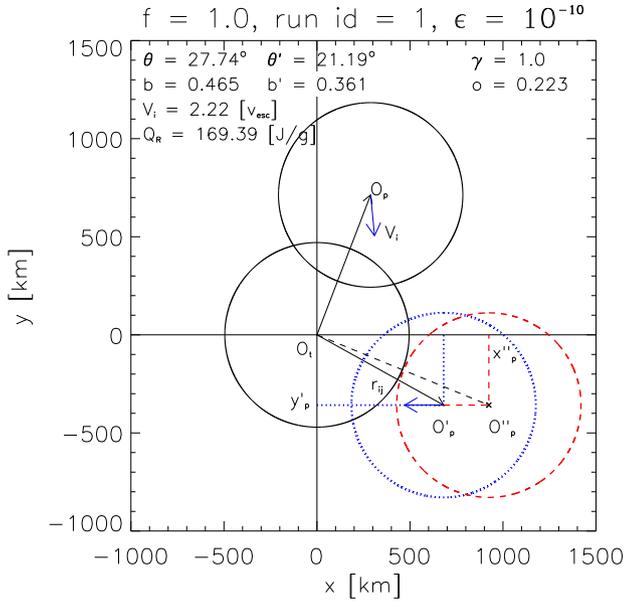}
\caption{A collision of equally massive bodies from the simulation 
with $f = 1$ and run id = 1 and $\epsilon = 10^{-10}$ is depicted. 
For details see the text.}
\label{Fig_06}
\end{figure}

\par
In Fig. \ref{Fig_06} a non-grazing collision from the simulation 
with $f = 1$, run id = 1 and $\epsilon = 10^{-10}$ is plotted. 
The reference frame is fixed at the target whose centre is marked 
by $O_{\mathrm t}$, the $O_{\mathrm t}x$ and $O_{\mathrm t}y$ 
axes are parallel with the $Ox$ and $Oy$ axes, respectively. 
In the figure $O_{\mathrm p}$ marks the centre of the projectile. 
The solid circles denote the colliding bodies. The origin of the 
impact velocity vector $\mathbf{V}_i$ is at $O_{\mathrm p}$. To 
better visualize the impact geometry the system is rotated until 
the impact velocity vector became parallel with the $O_{\mathrm t}x$ 
axis and points towards left (as in Fig. \ref{Fig_05}). The rotated 
projectile is plotted by a dashed blue circle and its centre is 
denoted by $O'_{\mathrm p} = (x'_{\mathrm p},\, y'_{\mathrm p})$. 
The impact parameter $b$ and angle $\theta$ are
\begin{equation}
b = \frac{|y'_{\mathrm p}|}{r_{\mathrm{ij}}}, \qquad \theta = \arcsin(b),
\label{eq:raw_b}
\end{equation}
where $r_{\mathrm{ij}} = \left|{\mathbf r}_{\mathrm j} - 
{\mathbf r}_{\mathrm i}\right|$ is the mutual distance of the point masses,
and ${\mathbf r}_{\mathrm i}$ is the particle's barycentric position vector.

\par
The projectile and the target are overlapping which is a consequence 
of the collision detection method applied in the numerical code. 
After each integration step the mutual distances of the bodies 
are calculated and compared against the sum of the bodies' radii 
multiplied by $f$. A collision happens whenever the
\begin{equation}
r_{\mathrm{ij}} \le f(R_{\mathrm i} + R_{\mathrm j})
\label{eq:collcrit}
\end{equation}
criterion fulfilled.

\par
When solving the equations of motion the particles are treated as 
point masses therefore two bodies can get arbitrary close. Since 
the collision detection is done after the integration step the 
collision can be detected only after the real physical contact 
hence overlapping is inevitable. The extent of the overlap can be 
defined as 
\begin{equation}
o = 1 - \frac{r_{\mathrm{ij}}}{f(R_{\mathrm i} + R_{\mathrm j})},
\label{eq:overlap}
\end{equation}
when $o=0$ the two spheres just touches each other. A consequence 
of the overlap is that the parameters defining the collisions cannot 
exactly determined from the data produced by the numerical simulation.
An improvement to this problem is described in Section 
\ref{section_improve_the_parameters}.

\par
\begin{figure}
\includegraphics[width=0.95\columnwidth,keepaspectratio=true]{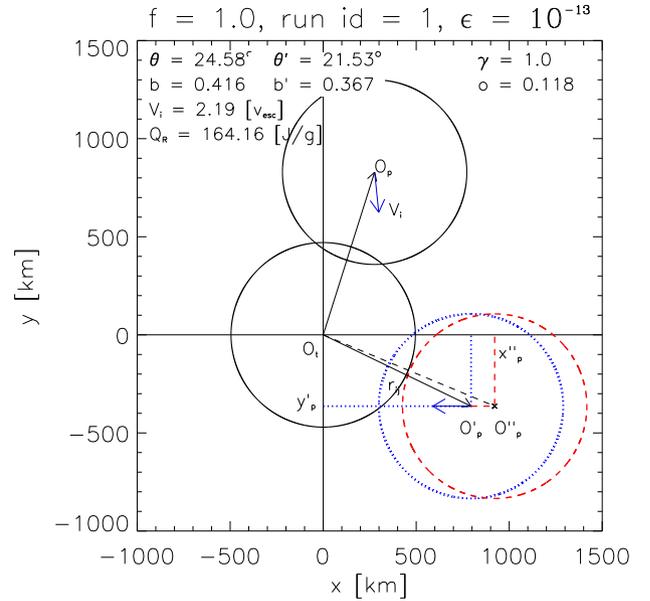}
\caption{Same as Fig. \ref{Fig_06} but for $\epsilon = 10^{-13}$.}
\label{Fig_07}
\end{figure}

The impact velocity is given in mutual escape velocity unit which 
is defined as follows \citep{Genda2012}:
\begin{equation}
V_{\mathrm{esc}} = k_\mathrm{G}\sqrt{2\frac{m_{\mathrm t} + m_{\mathrm p}}{R_{\mathrm t} + R_{\mathrm p}}},
\label{Eq_mutualescspeed}
\end{equation}
where $k_\mathrm{G}$ is the Gaussian constant of gravity. In the literature on planetary collisions $Q_{\mathrm R}$ denotes the specific energy of impact and it is given by
\begin{equation}
Q_{\mathrm R} = \frac{1}{2} \frac{m_{\mathrm t}v^2_{\mathrm t} + m_{\mathrm p}v^2_{\mathrm p}}{m_{\mathrm t} + m_{\mathrm p}},
\label{Eq_specificenergy}
\end{equation}
where $v_{\mathrm t}$ and $v_{\mathrm p}$ are the speed of the target and projectile with respect to the centre of mass, respectively. The specific impact energy is shown in the upper left corner of Figs. \ref{Fig_06} and \ref{Fig_07}. The projectile-to-target mass ratio, $\gamma = m_{\mathrm p}/m_{\mathrm t}$ and the overlap $o$ are displayed in the upper right corner. 

\subsection{Improve the parameters}
\label{section_improve_the_parameters}

The difficulty caused by the overlap of the bodies can be overcome by the 
following method. Let us assume that the direction of the relative 
velocity vector ${\mathbf v}_{\mathrm{ij}}$ does not change during 
the last integration step (after which the collision was detected). 
This assumption was checked with additional simulation where
the step size was 10 seconds and ${\mathbf v}_{\mathrm{ij}}$
was monitored. It was found that ${\mathbf v}_{\mathrm{ij}}$
hardly changed after the first contact.
In this case it is possible to shift the rotated projectile (dotted 
blue circle on Fig. \ref{Fig_06}) along the $x$-axis (i.e. parallel 
with $\mathbf{V}_{\mathrm i}$) until the first contact of projectile 
and target. The $x''_{\mathrm p}$-coordinate of the shifted projectile 
is (see Figs. \ref{Fig_06} and \ref{Fig_07})
\begin{equation}
x''_{\mathrm p} = \sqrt{f^2\left(R_{\mathrm t} + R_{\mathrm p}\right)^2 - y'^2_{\mathrm p}},
\label{Eq_xdp}
\end{equation}
and the improved collision parameter $b'$ and the associated collision 
angle $\theta'$ can be computed as
\begin{equation}
b' = \frac{|y'_{\mathrm p}|}{f\left(R_{\mathrm t} + R_{\mathrm p}\right)},\qquad \theta' = \arcsin (b').
\label{Eq_bprime}
\end{equation}
In Fig. \ref{Fig_06} one can see that $b = 0.465$ and $\theta = 27.74^\circ$, 
while the numeric values of the improved parameter is $b' = 0.361$ and 
$\theta' = 21.19^\circ$. Since $r_{\mathrm{ij}} \le f(R_{\mathrm t} + R_{\mathrm p})$ 
therefore it follows that $b' \le b \le 1$ and $\theta' \le \theta$ 
which means that the real impacts tend to be more head-on collisions 
than the ones detected from the raw numerical data.

\par
To check this method the results from the $\epsilon = 10^{-13}$ simulation 
is used and the same collision is displayed in Fig. \ref{Fig_07}. As 
expected by using higher accuracy the collision detection happens with 
a significantly  smaller overlap of 0.118. The computed $b = 0.416$ 
is smaller than in the $\epsilon = 10^{-10}$ case. From Fig. \ref{Fig_07} 
the improved parameter $b' = 0.367$, which agrees very well with the 
value computed from the $\epsilon = 10^{-10}$ case using Eq. (\ref{Eq_bprime}) 
proving the efficiency of the method.

\par
Using the above procedure and applying the law of energy conservation, the impact
velocity and specific impact energy can also improve. The total mechanical 
energy $E_1$ when the collision is detected (blue circles on Figs. \ref{Fig_06} 
and \ref{Fig_07}) and the energy of the shifted projectile (red circles) $E_2$
are equal and it allows one to compute the improved impact speed of the 
shifted projectile:
\begin{equation}
V_{\mathrm{i}}'=\sqrt{V_{\mathrm{i}}^2 - k_\mathrm{G}^2 m_{\mathrm t}\left(\frac{1}{r_{\mathrm ij}} - \frac{1}{f(R_i + R_j)} \right)}.
\label{Eq_Vi_prime}
\end{equation}

\par
Table \ref{Tab_03} presents the parameters discussed in detail
above for the two accuracy parameters. The value of $o$ decreased 
by 47\%, $b$ by 10\% and $b'$ increased by 2\% for the higher precision.
The $V_{\mathrm{i}}\,, V_{\mathrm{i}}'\,,Q_{\mathrm{R}}$ and 
$Q_{\mathrm{R}}'$ are almost the same for the two precisions, the 
relative changes are less than 3\%. From Eq. (\ref{Eq_Vi_prime}) 
it follows that the real impact velocity is slightly less than 
the measured one.

\begin{table}
\centering
\caption{Summary of parameters with their values from Figs. \ref{Fig_06} 
and \ref{Fig_07}. In the last two columns the absolute $\Delta$ and 
relative change $\Delta_{\mathrm r}$ is given where the reference 
values are taken from the $\epsilon = 10^{-10}$ case.}
\begin{tabular}{rrrr|rr}
     & $\epsilon = 10^{-10}$& $\epsilon = 10^{-13}$ & unit & $\Delta$ & $\Delta_{\mathrm r}$ [\%] \\
\hline
\hline
$o$               &   0.223 &   0.118 & - & -0.105 & 47.1 \\
$b$               &   0.465 &   0.416 & - & -0.049 & 10.5 \\
$b'$              &   0.361 &   0.367 & - &  0.006 &  1.7 \\
$V_{\mathrm{i}}$  &   2.222 &   2.187 & [$V_{\mathrm{esc}}$] & -0.035 & 1.6 \\
$V_{\mathrm{i}}'$ &   2.206 &   2.180 & [$V_{\mathrm{esc}}$] & -0.026 & 1.2 \\
$Q_{\mathrm{R}}$  & 169.39  & 164.16  & [J g$^{-1}$] & -5.24 & 3.1 \\
$Q_{\mathrm{R}}'$ & 166.92  & 163.01  & [J g$^{-1}$] & -3.91 & 2.3 \\
$\gamma$          &   1.0   &   1.0   & - & 0 & 0 \\
\hline
\end{tabular}
\label{Tab_03}
\end{table}

\subsection{Description of the collision data}
\label{section_desc_of_data}

During the simulations the $\mathbf{r_{\mathrm{i}}}$ and $\mathbf{v_{\mathrm{i}}}$ 
vectors and the physical properties of the colliding bodies were recorded in a 
database. Hereafter I refer to a dataset as {\it single run data} when I select 
from the database records with a specific $f$ and a single run id, and to a 
dataset as {\it multiple run data} when I select from the database records 
with a specific $f$ and multiple run ids. For the multiple run data I always 
selected the run ids from 1 to 10.

\par
In general at the end of each simulation ${\sim}10$ bodies remained so for 
each single run the number of collisions $n$, i.e. the sample size is 
$n \approx 9990$ therefore a multiple run data set contains approximately 
99 900 collisions. In total for the 50 runs there were 499507 and 499455 
collisions for $\epsilon = 10^{-10}$ and $\epsilon = 10^{-13}$, respectively. 
In this paper I used this database to study the distribution of the collision 
parameters. The details of the samples are given in Table \ref{Tab_A1} for 
$\epsilon = 10^{-10}$ and $\epsilon = 10^{-13}$, where the rows with 
run = 1 - 10 correspond to multiple run data, while the others correspond 
to single run data. In the table the minimum and maximum values of 
$V_{\mathrm{i}}$ and $Q_{\mathrm{R}}$ are also listed. For $\epsilon = 10^{-10}$ 
the values are on the left side, while for $\epsilon = 10^{-13}$ on the right side.

\par
According to the results the minimum value of $b'$ is $9.22\times 10^{-7}$ 
and $2.99\times 10^{-8}$ for $\epsilon = 10^{-10}$ and $\epsilon = 10^{-13}$, 
respectively, while the maximum is 1 for both $\epsilon$. This parameter 
has a theoretical range of [0, 1]. From the data it is clear that it does 
not depend on $f$. To save space $b'$ is not listed.

\par
In Table \ref{Tab_A1} with $f = 1$ the smallest value of $V_{\mathrm{i}}'$ 
is 0.978 while the biggest is 19.517. The variation of min($V_{\mathrm{i}}'$) 
is minor for a fixed $f$ and its value decreases with increasing $f$ 
from 0.99 to 0.2. The values of max($V_{\mathrm{i}}'$) do not decrease 
with increasing $f$, they are $8.356\substack{+2.001 \\ -1.146}$ except 
for one case for $f = 1$ and run id = 6 when max($V_{\mathrm{i}}'$) = 19.517. 
In Table \ref{Tab_A1} the data for this specific run without the 19.517 
value is also presented (empty cells have the same value as above). It 
turns out, that the maximum of $V_{\mathrm{i}}'$ drops down to 8.299, 
which is well in the other runs' range. This is an extreme value and 
in this single run data all other impacts are within the above range.

\par
The minimum of $Q_{\mathrm{R}}'$ is 1.23$\times 10^{4}$ J kg$^{-1}$ while 
the maximum is 1.31$\times 10^{7}$ J kg$^{-1}$ for $f = 1$. The minimum 
values fall in a rather narrow range for a specific $f$ which decreases 
with increasing $f$. The maximum values fall in a limited range for a 
specific $f$ which shows a decreasing trend with increasing $f$.

\par
For each $f$ the table shows the range for the multiple run data, 
designated by 1 - 10 in the run column. For $f = 1$ the data are 
listed containing the extreme impact speed and without it.

\par
As reported by Table \ref{Tab_A1} in the case of $\epsilon = 10^{-13}$ 
the sample sizes are in almost perfect agreement with the lower accuracy
runs, the ranges of the parameters are very similar and the data show 
analogous behaviour. In this case there is no extreme value. Comparing 
the results for the two accuracy parameters one can conclude that the 
size of the populations, the ranges of the impact speed and specific 
impact energy are very similar. 

\par
In Table \ref{Tab_04} the mean of the minimum and maximum values of 
$b'$ are calculated and listed along with their standard deviations. 
It is clear that these values practically do not depend on $f$: the 
minimum is very close to 0 while the maximum to 1, the standard 
deviation is very limited for both quantities, the deviations around 
the mean is small. The figures are alike for $\epsilon = 10^{-13}$, 
therefore they are not shown.

\par
In Table \ref{Tab_05} the mean of the minimum and maximum values 
of $V_{\mathrm{i}}'$ are calculated from the values given in Table 
\ref{Tab_A1} and listed along with their standard deviations for 
$\epsilon = 10^{-10}$ and $\epsilon = 10^{-13}$. The mean of the 
minimum value depends strongly on $f$, it decreases from 0.99 to 
0.28 in both cases, while the maximum essentially does not depend 
on it, but stays around 8.3. The standard deviation is confined 
to a narrow interval for both the minimum and maximum, indicating 
modest variability of the values.

\par
In Table \ref{Tab_06} the extreme values and the mean of $Q_{\mathrm{R}}'$ 
derived from the data given in Table \ref{Tab_A1} are shown along 
with their standard deviations for the two accuracy parameters. As 
expected, the mean of the minimum value depends strongly on $f$, it 
decreases from ${\sim}13 \times 10^{3}$ to ${\sim}1.8 \times 10^{3}$ in
both cases, while the maximum drops from ${\sim}4.4 \times 10^{6}$ 
to ${\sim}2.7 \times 10^{6}$ and than stays constant around this 
value as $f$ increases.

\par
As a final note the mean, median and sd for all multiple run data 
were calculated for the two accuracy parameters. These results are 
presented in Table \ref{Tab_07}. The median is the value separating 
the higher half from the lower half of a data sample. The impact 
parameters' mean and median value are close to 0.5, which is the 
middle of its range. In the case of $V_{\mathrm{i}}'$ the mean is 
definitely larger than the median for all data sets. This implies 
that there are some large values which affect the mean and most of 
the values are closer to the median than to the mean. This fact 
also reflects in the relative large sd values. Since the specific 
impact energy is a strong function of the impact speed, the median 
of $Q_{\mathrm{R}}'$ is significantly smaller than the mean. Again 
this is visible from the larger spread of the data, given the 
larger sd values.

\begin{table}
\centering
\caption{The mean of the minimum (\nth{2} column) and maximum (\nth{4} column) 
of $b'$ for multiple run data and their standard deviations in the \nth{3} 
and \nth{5} columns, respectively.}
\begin{tabular}{rrrrr}
$f$ & $\overline{\mbox{min}(b')}$ & sd(min($b'$)) & $\overline{\mbox{max}(b')}$ & sd(max($b'$)) \\
\hline
\hline
 1 & 1.56$\times 10^{-4}$ & 1.28$\times 10^{-4}$ & 0.999985 & 4.74$\times 10^{-5}$ \\
 2 & 6.80$\times 10^{-5}$ & 5.84$\times 10^{-5}$ & 0.999686 & 3.88$\times 10^{-4}$ \\
 3 & 1.22$\times 10^{-4}$ & 9.77$\times 10^{-5}$ & 0.999404 & 6.96$\times 10^{-4}$ \\
 5 & 5.94$\times 10^{-5}$ & 6.32$\times 10^{-5}$ & 0.999093 & 7.41$\times 10^{-4}$ \\
10 & 1.40$\times 10^{-4}$ & 8.73$\times 10^{-5}$ & 0.999274 & 4.04$\times 10^{-4}$ \\
\hline
\end{tabular}
\label{Tab_04}
\end{table}

\begin{table}
\centering
\caption{The same for $V'_\mathrm{i}$ as in Table \ref{Tab_04}.}
\begin{tabular}{rrrrr}
$f$ & $\overline{\mbox{min}(V_\mathrm{i}')}$ & sd(min($V_\mathrm{i}'$)) & $\overline{\mbox{max}(V_\mathrm{i}')}$ & sd(max($V_\mathrm{i}'$)) \\
\hline
\multicolumn{5}{c}{$\epsilon = 10^{-10}$} \\
\hline
 1 &  0.991 &  0.005 &  8.712 &  0.805 \\
 2 &  0.689 &  0.016 &  8.268 &  0.587 \\
 3 &  0.559 &  0.008 &  8.119 &  0.517 \\
 5 &  0.418 &  0.016 &  8.401 &  0.743 \\
10 &  0.275 &  0.029 &  8.278 &  0.484 \\
\hline
\multicolumn{5}{c}{$\epsilon = 10^{-13}$} \\
\hline
 1 &  0.990 &  0.004 &  8.322 &  0.641 \\
 2 &  0.686 &  0.020 &  8.260 &  0.789 \\
 3 &  0.554 &  0.015 &  8.062 &  0.779 \\
 5 &  0.423 &  0.009 &  8.134 &  0.643 \\
10 &  0.279 &  0.015 &  8.105 &  0.520 \\
\hline
\end{tabular}
\label{Tab_05}
\end{table}

\begin{table}
\centering
\caption{The same for $Q_\mathrm{R}'$ as in Table \ref{Tab_04}.}
\begin{tabular}{rrrrr}
$f$ & $\overline{\mbox{min}(Q_\mathrm{R}')}$ & sd(min($Q_\mathrm{R}'$)) & $\overline{\mbox{max}(Q_\mathrm{R}')}$ & sd(max($Q_\mathrm{R}'$)) \\
\hline
\multicolumn{5}{c}{$\epsilon = 10^{-10}$} \\
\hline
 1 & 1.312$\times 10^{4}$ & 1.485$\times 10^{3}$ & 4.415$\times 10^{6}$ &  7.681$\times 10^{5}$ \\
 2 & 7.670$\times 10^{3}$ & 7.084$\times 10^{2}$ & 2.665$\times 10^{6}$ &  5.158$\times 10^{5}$ \\
 3 & 5.473$\times 10^{3}$ & 7.611$\times 10^{2}$ & 2.378$\times 10^{6}$ &  3.368$\times 10^{5}$ \\
 5 & 3.489$\times 10^{3}$ & 3.670$\times 10^{2}$ & 2.454$\times 10^{6}$ &  4.300$\times 10^{5}$ \\
10 & 1.789$\times 10^{3}$ & 2.761$\times 10^{2}$ & 2.358$\times 10^{6}$ &  2.778$\times 10^{5}$ \\
\hline
\multicolumn{5}{c}{$\epsilon = 10^{-13}$} \\
\hline
 1 & 1.279$\times 10^{4}$ & 1.756$\times 10^{3}$ & 4.498$\times 10^{6}$ &  1.125$\times 10^{6}$ \\
 2 & 7.127$\times 10^{3}$ & 5.111$\times 10^{2}$ & 2.957$\times 10^{6}$ &  8.200$\times 10^{5}$ \\
 3 & 5.123$\times 10^{3}$ & 6.030$\times 10^{2}$ & 2.459$\times 10^{6}$ &  6.417$\times 10^{5}$ \\
 5 & 3.030$\times 10^{3}$ & 4.610$\times 10^{2}$ & 2.282$\times 10^{6}$ &  3.684$\times 10^{5}$ \\
10 & 1.776$\times 10^{3}$ & 3.311$\times 10^{2}$ & 2.262$\times 10^{6}$ &  2.803$\times 10^{5}$ \\
\hline
\end{tabular}
\label{Tab_06}
\end{table}

\begin{table*}
\centering
\caption{The mean, median and standard deviation of $b'$, $V_{\mathrm{i}}'$ 
and $Q_{\mathrm{R}}'$. The results are given for all multiple run data.}
\begin{tabular}{rccccccccc}
\hline  
$f$ & $\overline{b'}$ & Med($b'$) & sd($b'$) & $\overline{V_{\mathrm{i}}'}$ & Med($V_{\mathrm{i}}'$) & sd($V_{\mathrm{i}}'$) & $\overline{Q_{\mathrm{R}}'}$ & Med($Q_{\mathrm{R}}'$) & sd($Q_{\mathrm{R}}'$) \\
& & & & & & & $\times 10^{5}$ & $\times 10^{5}$ & $\times 10^{5}$ \\
\hline
\multicolumn{10}{c}{$\epsilon = 10^{-10}$} \\ 
\hline
 1 & 0.486 & 0.482 & 0.283 & 1.836 & 1.517 & 0.903 & 1.964 & 1.311 & 2.184 \\
 2 & 0.487 & 0.485 & 0.284 & 1.627 & 1.309 & 0.961 & 1.541 & 0.941 & 1.777 \\
 3 & 0.489 & 0.488 & 0.284 & 1.550 & 1.234 & 0.991 & 1.406 & 0.804 & 1.707 \\
 5 & 0.486 & 0.483 & 0.282 & 1.479 & 1.173 & 1.023 & 1.298 & 0.686 & 1.686 \\
10 & 0.485 & 0.483 & 0.282 & 1.419 & 1.129 & 1.051 & 1.216 & 0.600 & 1.673 \\
\hline
\multicolumn{10}{c}{$\epsilon = 10^{-13}$} \\ 
\hline
 1 & 0.494 & 0.492 & 0.286 & 1.826 & 1.508 & 0.907 & 1.947 & 1.286 & 2.159 \\
 2 & 0.494 & 0.491 & 0.286 & 1.624 & 1.302 & 0.969 & 1.535 & 0.927 & 1.782 \\
 3 & 0.495 & 0.494 & 0.287 & 1.544 & 1.224 & 0.998 & 1.402 & 0.793 & 1.721 \\
 5 & 0.495 & 0.495 & 0.287 & 1.480 & 1.173 & 1.033 & 1.304 & 0.685 & 1.710 \\
10 & 0.493 & 0.491 & 0.287 & 1.424 & 1.131 & 1.062 & 1.229 & 0.599 & 1.699 \\
\hline
\end{tabular}
\label{Tab_07}
\end{table*}

\subsection{The bin size}
\label{section_bin_size}

To construct the histograms, the first step is to bin the range of values 
and then count how many values fall into each interval. In this paper the 
bins are consecutive, adjacent and non-overlapping intervals of a variable. 
The bins have equal size. Since there is no ultimate method to determine 
the number of bins, after some experimentation I found that the Rice rule 
gives the best number of bins:
\begin{equation}
k = [2n^\frac{1}{3}], \qquad bs = \frac{\mathrm{max} - \mathrm{min}}{k},
\label{eq:rice-rule}
\end{equation}
where $bs$ is the bin size, $\mathrm{min},\,\mathrm{max}$ denotes the minimum 
and maximum of the underlying variable and the braces indicate the ceiling 
function. Throughout this paper I use normalized histograms to display 
relative frequencies which is the proportion of cases that fall into 
each of several categories, with the sum of the heights equalling 1. 
Histograms give an approximation of the probability density function, 
pdf of the underlying variable. 

\par
A cumulative histogram refers to the running total of the values. 
That is, the cumulative histogram $H_i$ of a histogram $h_j$ is defined as:
\begin{equation}
H_i = \sum_{j=1}^{i}h_j.
\label{Eq_H_i}
\end{equation}
Since the collision parameters have continuous probability distribution 
hereafter I will use the terms pdf and cdf.

\subsection{Distribution of collision parameters}
\label{section_dist_coll_param}

\subsubsection{Analysis of the correction for the overlap}

\begin{figure*}
\includegraphics[width=0.94\columnwidth,keepaspectratio=true]{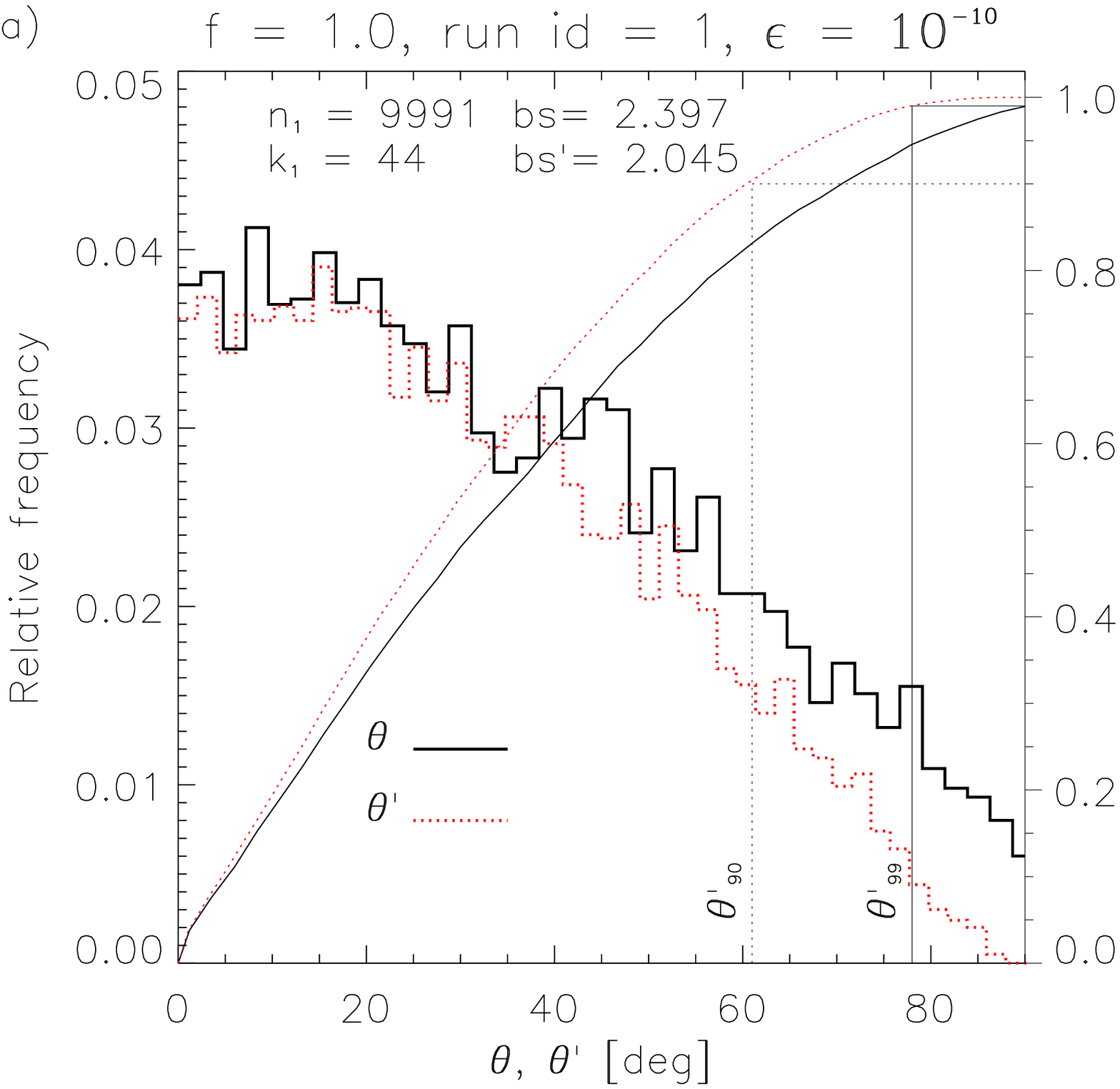}
\includegraphics[width=0.94\columnwidth,keepaspectratio=true]{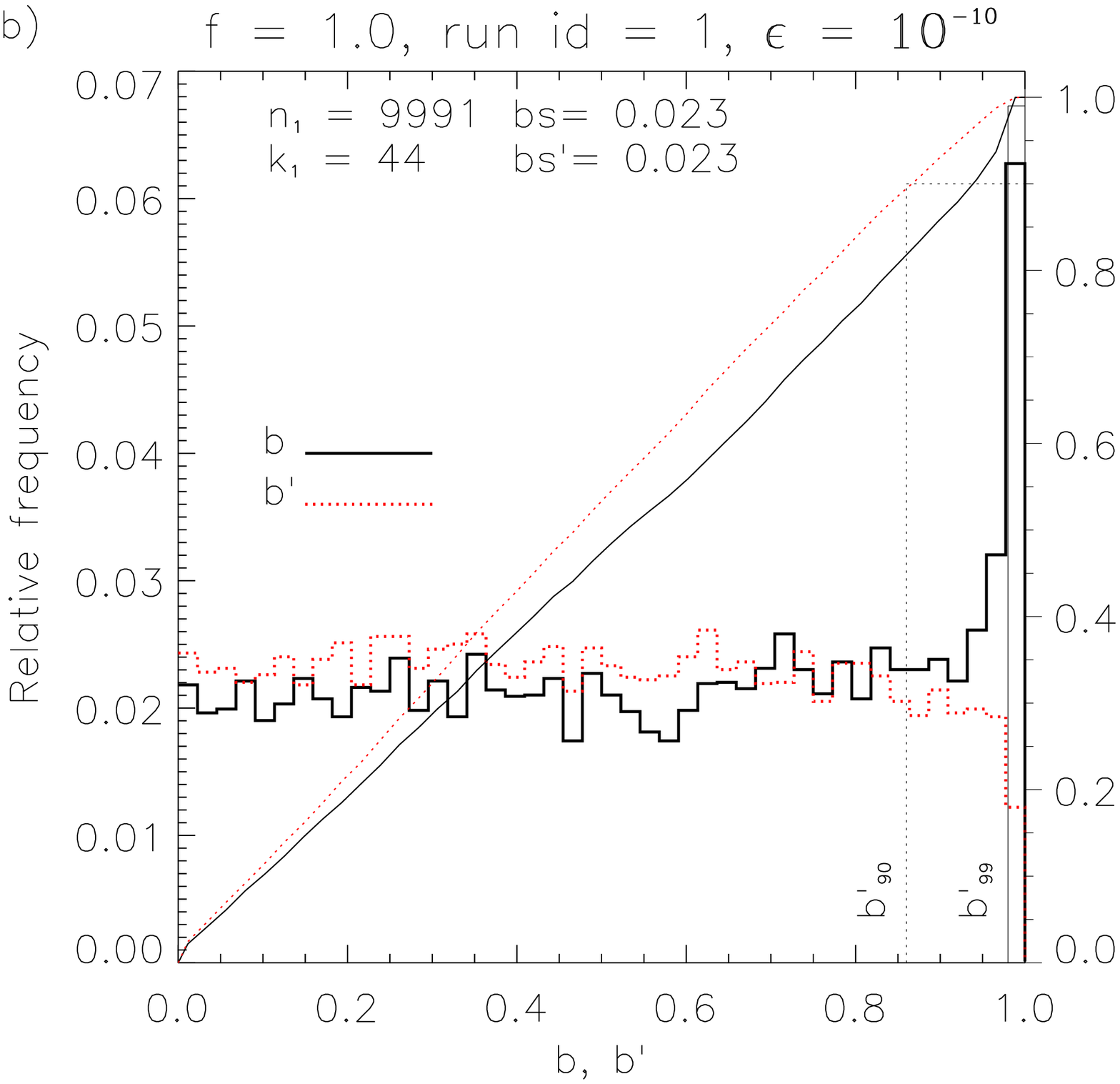}
\includegraphics[width=0.94\columnwidth,keepaspectratio=true]{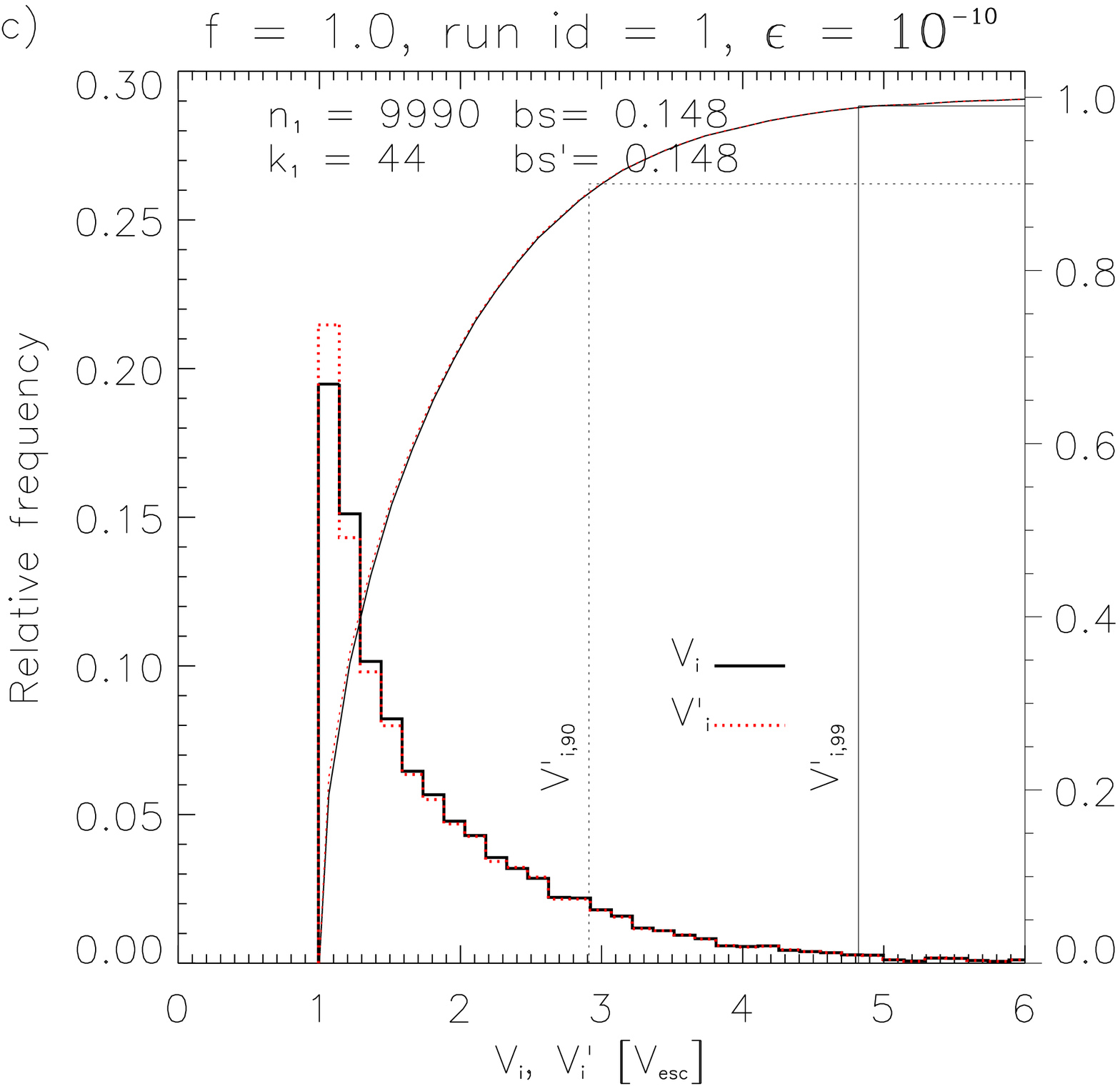}
\includegraphics[width=0.94\columnwidth,keepaspectratio=true]{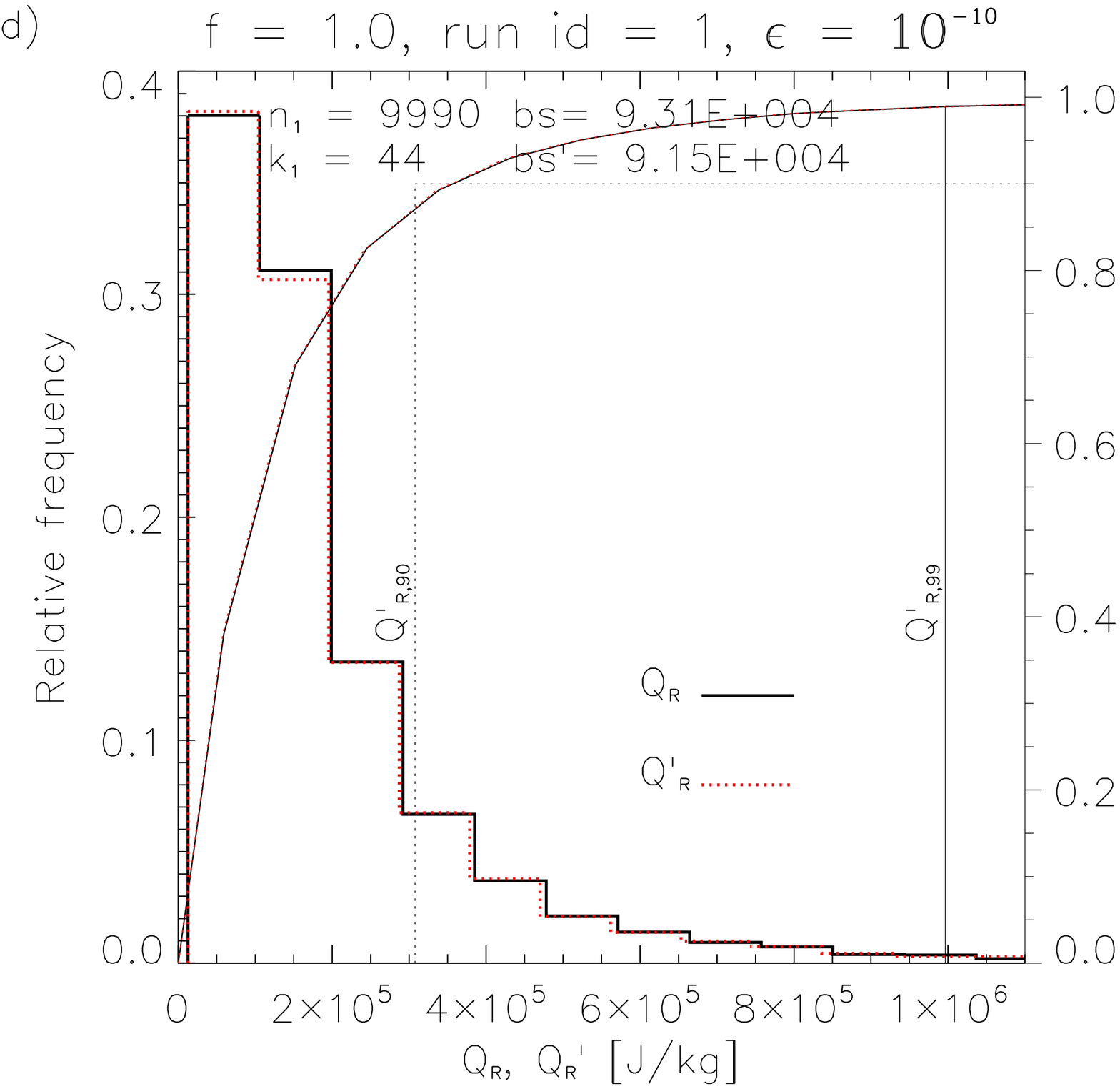}
\caption{Comparison of the distribution of the impact parameters with 
their improved version for $\epsilon = 10^{-10}$ and run id = 1. 
On the left vertical axis the relative frequency is shown and the 
corresponding pdf curves are plotted with solid black and dotted red line 
segments for the raw and the improved quantity, respectively. 
The cdf curves computed from Eq. \ref{Eq_H_i} for the raw and improved
quantities are also plotted with thin black and dotted red lines. 
The scale for the cdf is given on the right vertical axis.
On panel a) $\theta$ and $\theta '$ are compared. The two vertical lines 
denotes the location where cdf reaches 90\% and 99\%, respectively. On 
panel b) the $b$ and $b'$, on panel c) $V_{\mathrm{i}}$ and $V_{\mathrm{i}}'$ 
while on panel d) $Q_{\mathrm{R}}$ and $Q_{\mathrm{R}}'$ are plotted. The 
legends are the same for each panel.}
\label{Fig_08}
\end{figure*}

In section \ref{section_improve_the_parameters} a method
was presented to improve the impact parameter, speed 
and specific impact energy (see Eqs. (\ref{Eq_bprime})
and (\ref{Eq_Vi_prime})). Here I show how it 
influences their distribution. In Fig. \ref{Fig_08} 
I have summarized the result for $f = 1$ and 
run id = 1 with $\epsilon = 10^{-10}$. Panel a) shows 
the pdf of $\theta$ (solid black curve) and $\theta '$
(red dotted curve), panel b) displays the pdf of $b$
(solid black curve) and $b'$ (red dotted curve). The
number of collision is $n_1 = 9991$ and the bin size is
$bs(\theta') = 2.045^\circ$ and $bs(b) = bs(b') = 0.023$. 
The bin size for $\theta$ is $\approx 2.4^\circ$ which is
larger than for $\theta'$. The reason for this difference
originate from that $\mathrm{max}(\theta) = 105^\circ$ 
which is larger than the theoretically allowed $90^\circ$; 
another reason to improve the parameters.

In Fig. \ref{Fig_08} from panel a) it is clearly visible that the 
distribution of $\theta'$ is different from that of $\theta$. 
For $\theta < 40^\circ$ the relative frequency of $\theta$ is similar 
to that of $\theta'$, while for $\theta > 40^\circ$ the pdf curves 
diverge more and more, the relative frequency of $\theta$ 
is higher than $\theta'$. Our 2D results do not support the 
observation of \cite{Shoemaker1962}, who concludes that a $45^\circ$ 
impact angle ($b = 0.707$) is most probable. Preliminary results of 
3D runs shows evidence of a peak around $45^\circ$ therefore it seems 
that there is a significant difference between 2D and 3D model when 
collision parameter is an important aspect. The causes of this 
discrepancy should be be discussed in a separate paper.
From Table \ref{Tab_08} cdf($\theta '$) reaches 90\% at 
$\theta ' \approx 61^\circ$ and 99\% at $\theta ' \approx 77^\circ$. 
These values are shown with dotted and solid black vertical lines 
denoted by $\theta'_{90}$ and $\theta'_{99}$, respectively.

\begin{figure*}
\includegraphics[width=0.94\columnwidth,keepaspectratio=true]{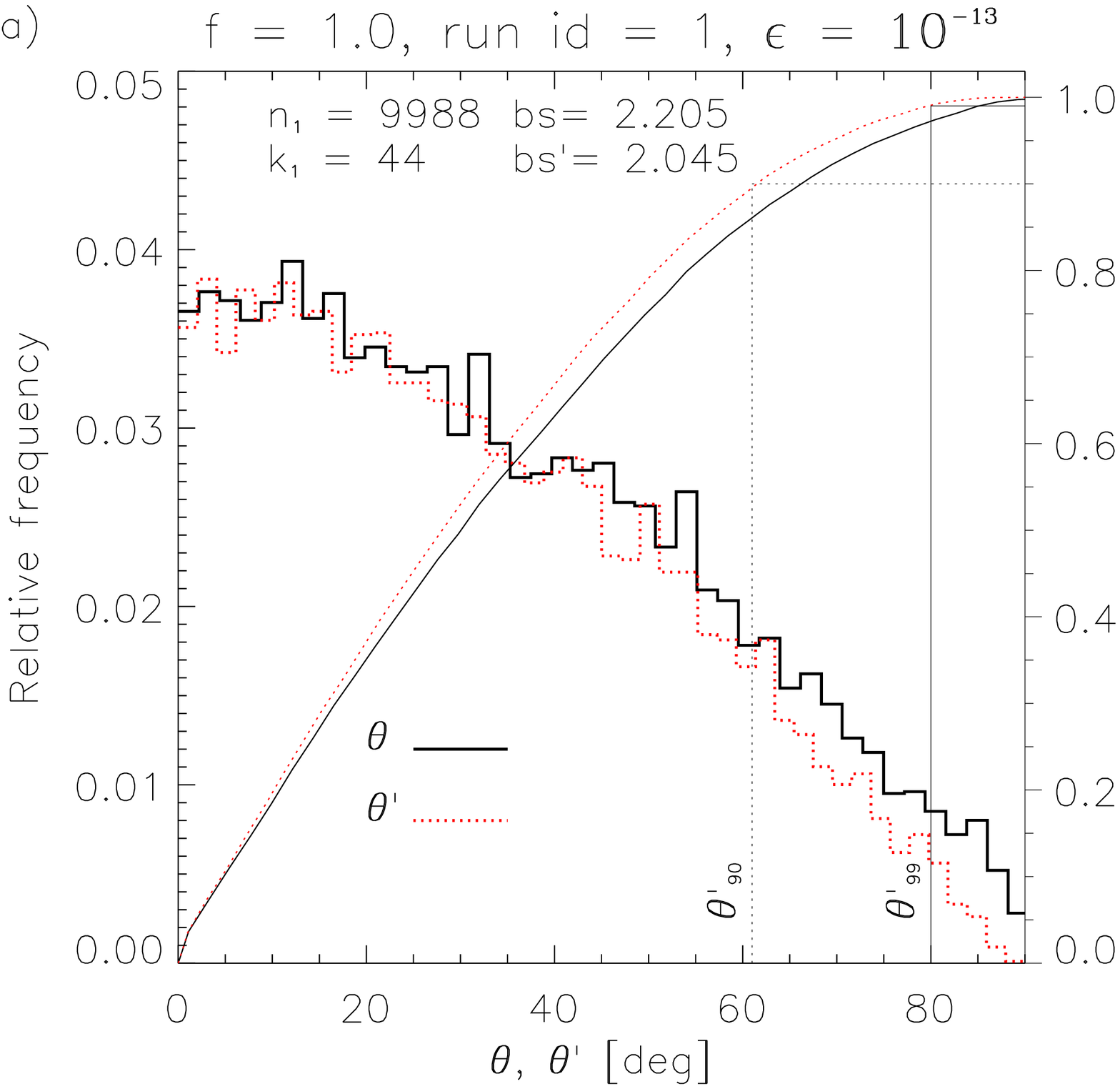}
\includegraphics[width=0.94\columnwidth,keepaspectratio=true]{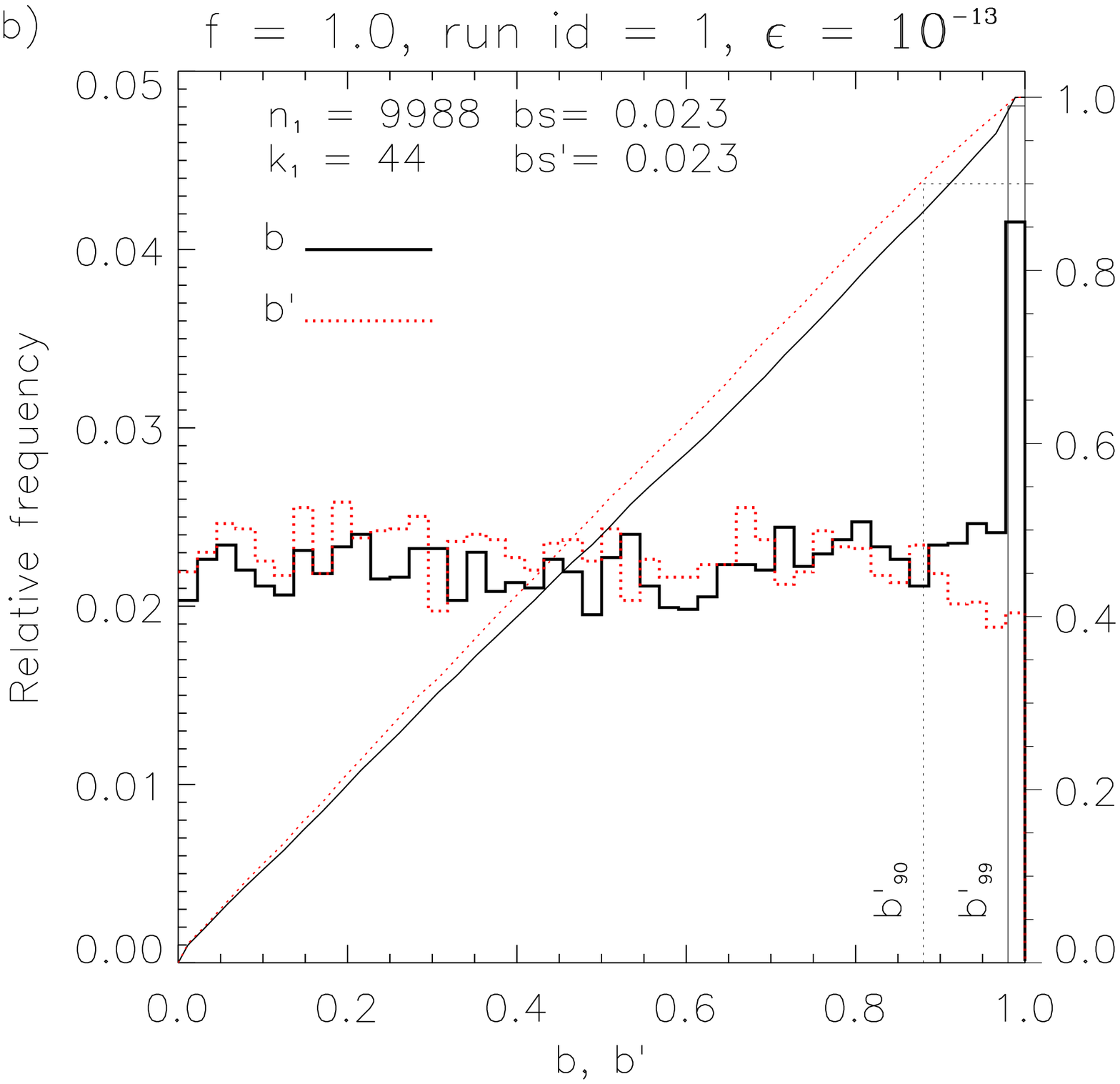}
\includegraphics[width=0.94\columnwidth,keepaspectratio=true]{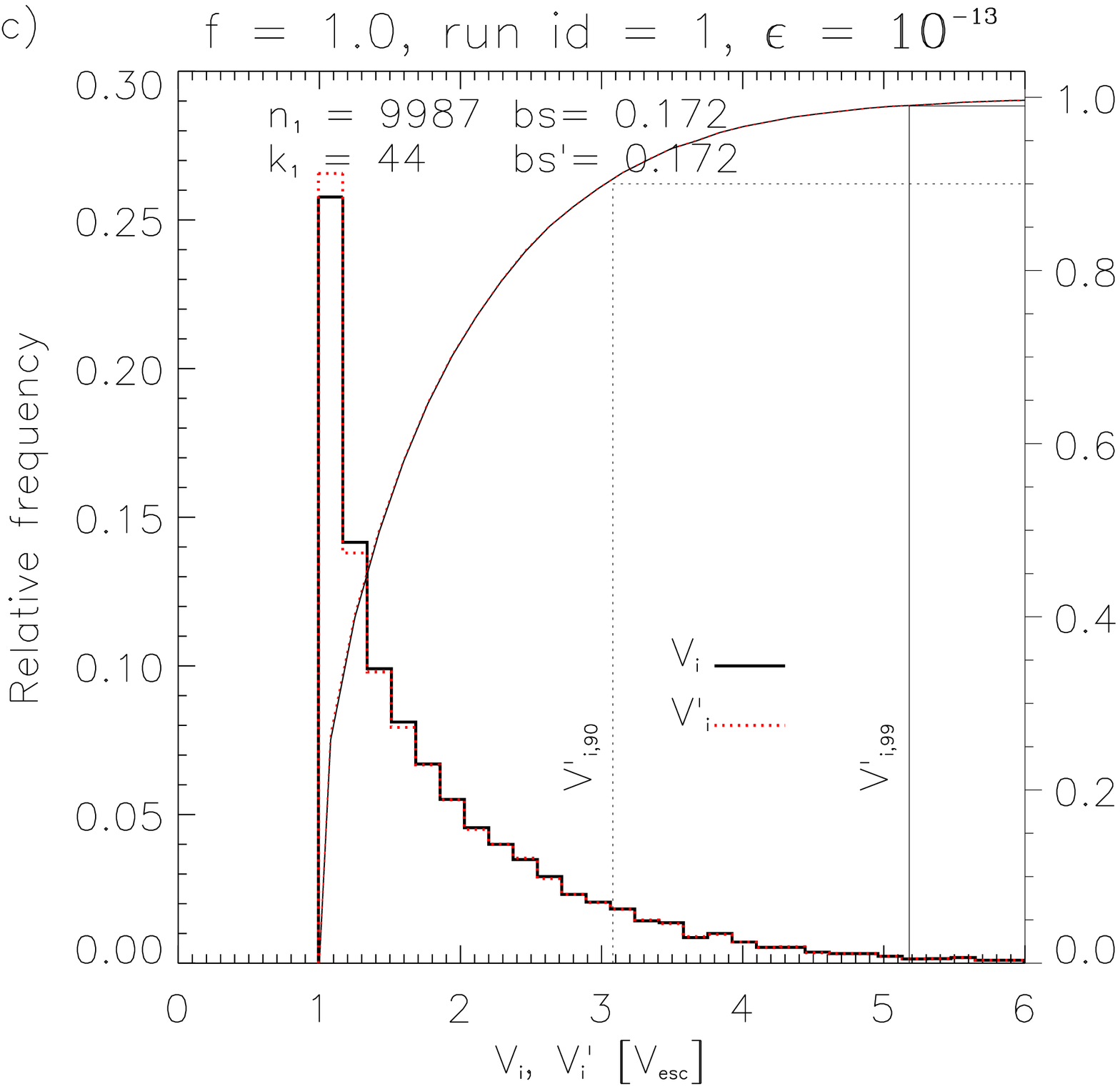}
\includegraphics[width=0.94\columnwidth,keepaspectratio=true]{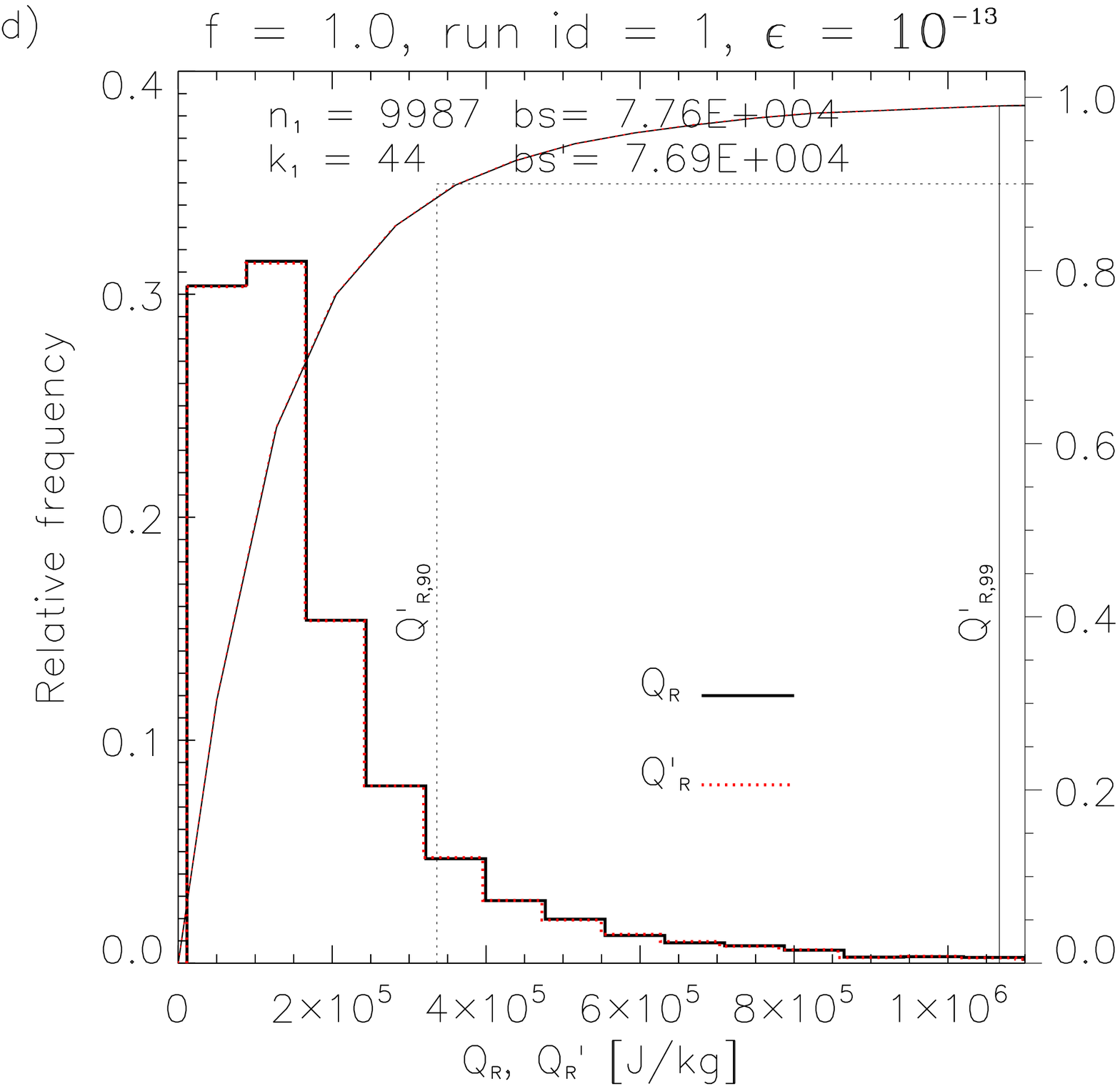}
\caption{The same as Fig. \ref{Fig_08} but for $\epsilon = 10^{-13}$.}
\label{Fig_09}
\end{figure*}

\par
In Fig. \ref{Fig_08} panel b) the distribution of $b$ and $b'$ are depicted. 
These are very similar for $b \le 0.8$ ($\approx 53^\circ$): the frequency 
of $b'$ is about 5 - 10\% higher than $b$. For $b > 0.9$ the difference 
becomes prominent. For the last bin the relative frequency of $b$ is 0.06 
while it is 0.01 for $b'$. From Table \ref{Tab_08} the cdf($b'$) reaches 
0.9 at $b' \approx 0.864$ and 0.99 at $b' \approx 0.977$. These values are 
shown with dotted and solid black vertical lines denoted by $b'_{90}$ and 
$b'_{99}$, respectively. By eyeballing the data it is reasonable to state 
that the distribution of $b'$ is uniform in the range [0, 0.95], and has a 
sharp decrease in the last bin.

\par
In Fig. \ref{Fig_08} panel c) the distributions of $V_{\mathrm{i}}$ and 
$V_{\mathrm{i}}'$ are shown. The pdf and cdf functions for these quantities 
are very similar. According to Table \ref{Tab_08} one gets that 90\% of 
$V_{\mathrm{i}}'$ are less than 2.908 and 99\% are less then 4.824. These 
values are shown with dotted and solid black vertical lines denoted by 
$V_{\mathrm{i,90}}'$ and $V_{\mathrm{i,99}}'$, respectively. There are 
no data with $V_{\mathrm{i}}' < 0.993$, the maximum of the pdf is around 
1 than it decreases fast as $V_{\mathrm{i}}'$ gets larger. For 
$V_{\mathrm{i}} \ge 5$ the probability of a collision is less than 1\%.

\par
Panel d) shows the distribution of $Q_{\mathrm{R}}$ and $Q_{\mathrm{R}}'$. 
The pdf and cdf functions are almost identical. From Table \ref{Tab_08} 
cdf reaches 90\% at $Q_{\mathrm{R}}' \approx 3.07 \times 10^{5}$ and 99\% 
at $Q_{\mathrm{R}}' \approx 10^{6}$ J kg$^{-1}$. These values are shown 
with dotted and solid black vertical lines denoted by $Q_{\mathrm{R,90}}'$ 
and $Q_{\mathrm{R,99}}'$, respectively.

\par
As it is apparent from Fig. \ref{Fig_08} the difference between the 
raw and improved quantities are not negligible, primarily in the case of 
the impact angle, parameter (panel a, b) and speed (panel c). This finding 
provides a further argument to use the correction for the overlap method 
described in section \ref{section_improve_the_parameters}.

\par
In Fig. \ref{Fig_09} the same quantities for $\epsilon = 10^{-13}$ are 
shown. The overall behavior of the curves are very similar to the 
$\epsilon = 10^{-10}$ case but, as expected, the difference between the 
raw and improved quantities are smaller.

\par
\begin{figure*}
\includegraphics[width=0.9\columnwidth,keepaspectratio=true]{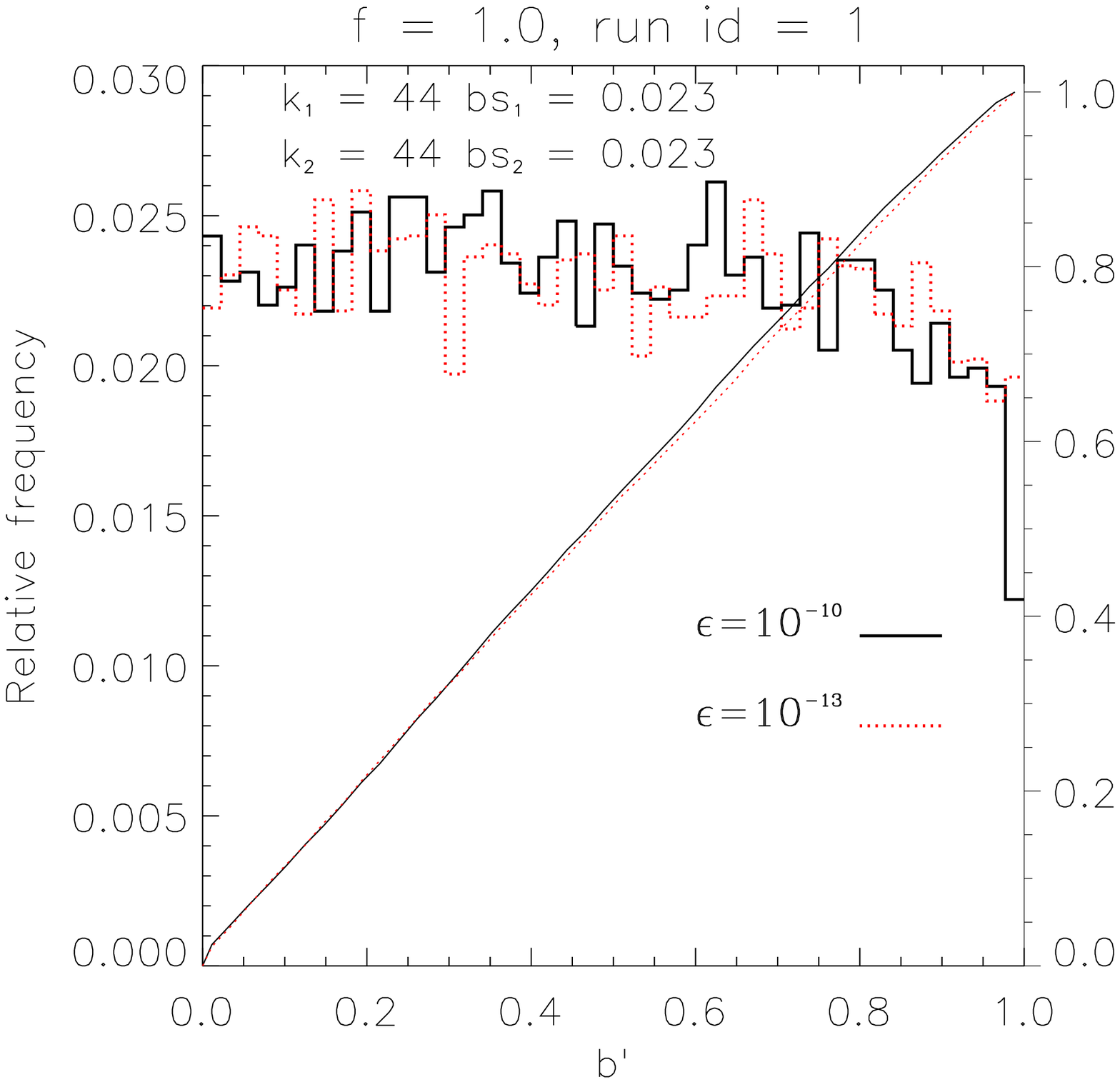}
\includegraphics[width=0.9\columnwidth,keepaspectratio=true]{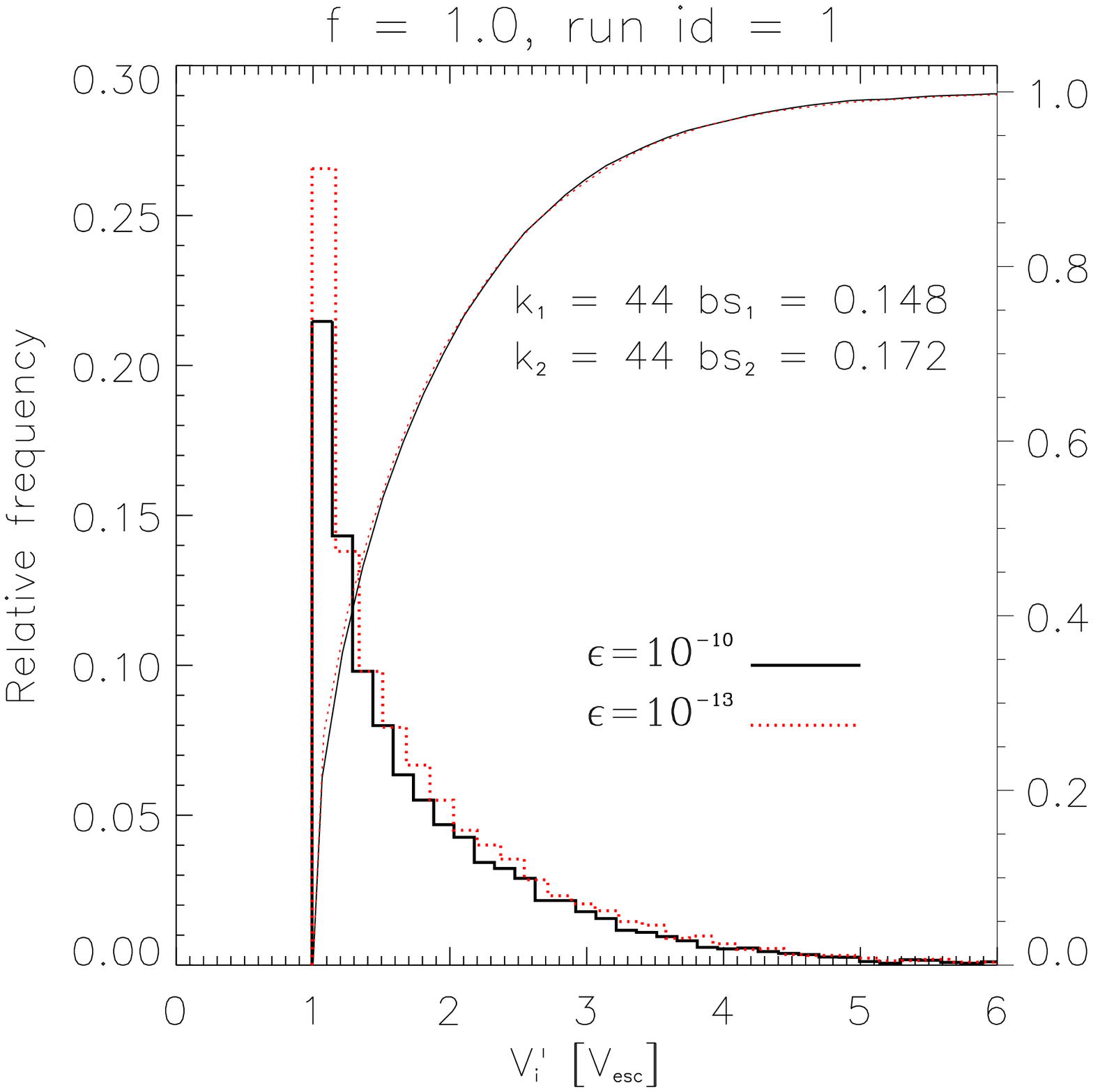}
\caption{On the left panel the distribution of $b'$, on the right panel $V_{\mathrm{i}}'$ is compared for $\epsilon = 10^{-10}$ and $\epsilon = 10^{-13}$.}
\label{Fig_10}
\end{figure*}

To compare directly the results produced by the runs with $\epsilon = 10^{-10}$ 
and $\epsilon = 10^{-13}$ in Fig. \ref{Fig_10} the distribution of $b'$ are 
presented on the left panel for both accuracy values. Apart from a random 
variation the pdf curves are akin and fluctuates around the same mean while 
the cdf curves practically cannot be distinguished from each other. From 
the right panel of Fig. \ref{Fig_10} there is apparently a notable difference 
between the pdf curves of $V_{\mathrm{i}}'$ around 1, where the red dotted 
curve is about 0.05 higher than the solid black one indicating more frequent 
collision with lower velocity when the accuracy is higher. Beyond 1 the 
shape of the two curves is very similar, although the dotted curve is 
slightly above the black curve, consequently there are more impact with 
lower velocity for the higher accuracy. The cdf curves are almost identical.

\par
From these comparisons and analysis it is evident that one should always 
correct the impact parameters as it was described in section 
\ref{section_improve_the_parameters} and from a statistical point of view 
it is enough to follow the evolution of the system with a lower accuracy 
parameter if one is interested only in the collision statistical properties 
of the system. The only exception is the frequency of collisions around 1, 
where the lower accuracy simulations provide lower frequency.

\begin{table}
\centering
\caption{Summary of the values for $f = 1$ and run id = 1 at which 
the cdf reaches 90\% and 99\% for $\epsilon = 10^{-10}$ and 
$\epsilon = 10^{-13}$.}
\begin{tabular}{rrr|rr}
\hline
 & \multicolumn{2}{c|}{$\epsilon = 10^{-10}$} & \multicolumn{2}{c}{$\epsilon = 10^{-13}$} \\
\hline
cdf     & 90\% & 99\% & 90\% & 99\% \\
\hline
$\theta'$ [deg]   &  61.364 &  77.727 & 61.364 &  79.773  \\
$b'$              &   0.864 &   0.977 &  0.886 &   0.977  \\
$V_{\mathrm{i}}'$ [$V_\mathrm{esc}$] &   2.908 &   4.824 & 3.084 & 5.176 \\
$Q_{\mathrm{R}}'$ [J kg$^{-1}$] & 307734 & 996513 & 335928 & 1066795 \\
\hline
\end{tabular}
\label{Tab_08}
\end{table}

\subsubsection{Comparison of statistics derived from single runs}

\begin{figure*}
\includegraphics[width=0.9\columnwidth,keepaspectratio=true]{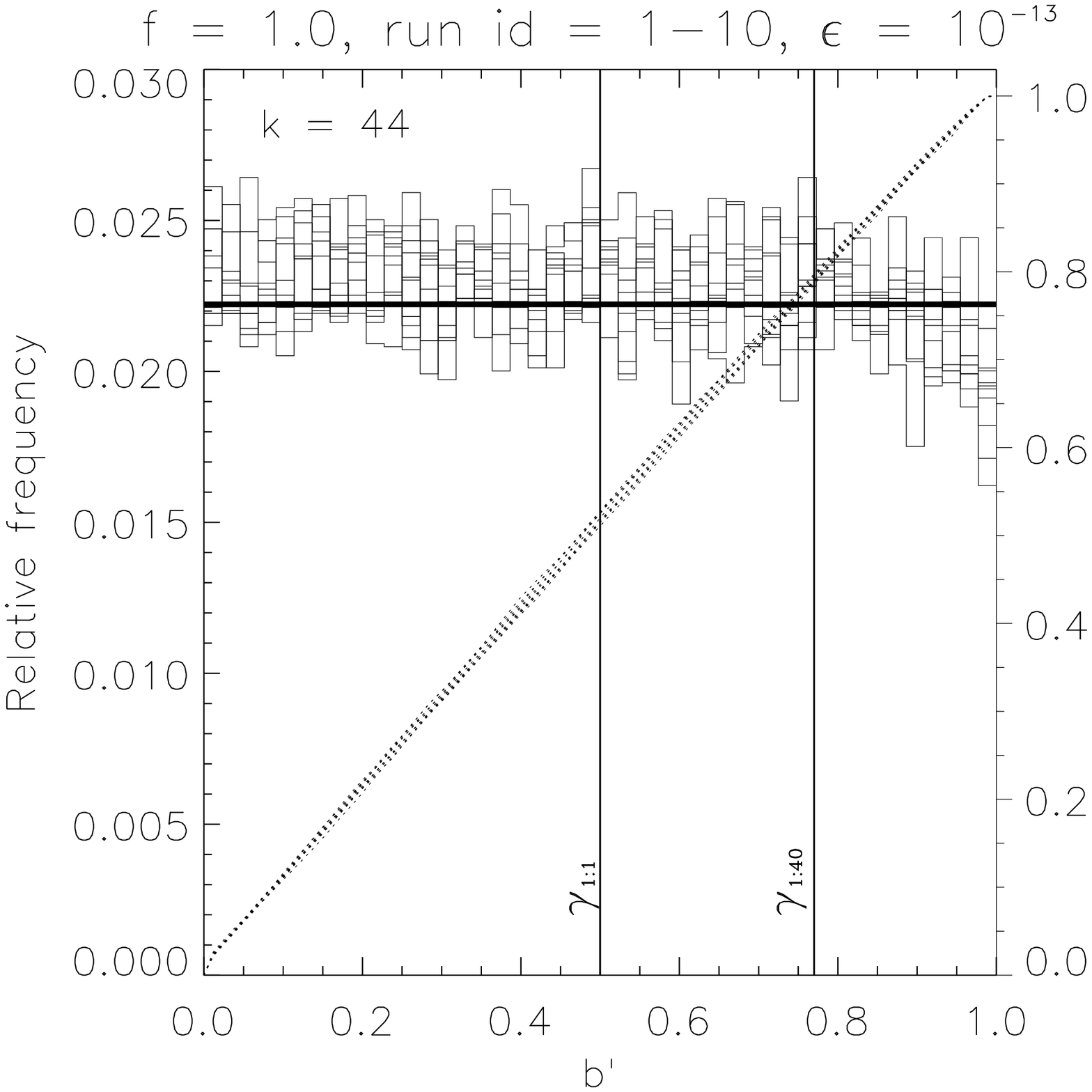}
\includegraphics[width=0.9\columnwidth,keepaspectratio=true]{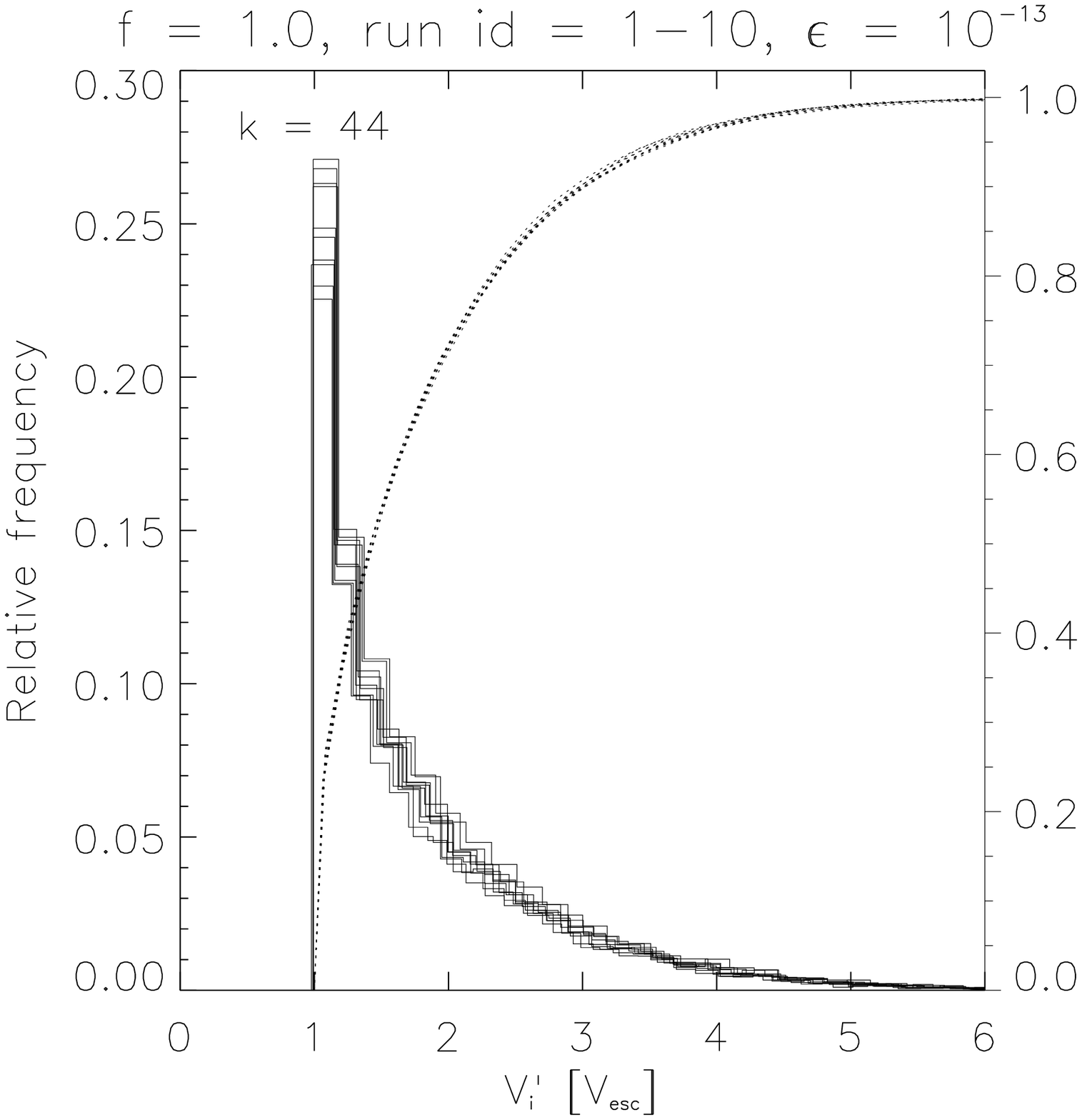}
\caption{Summary of the results for $\epsilon = 10^{-13}$ for 
run = $1,\ldots, 10$. On the left panel the distribution of $b'$ 
is displayed. The thick black horizontal line is the mean of all 
the 10 pdfs. The vertical lines denotes the threshold of grazing 
impacts between equal-sized bodies ($\gamma_{1:1}$) and between 
bodies with $\gamma = 1:40$. On the right panel the pdfs of 
$V_{\mathrm{i}}'$ is shown. For details see text.}
\label{Fig_11}
\end{figure*}

It is instructive to compare the results of the 10 different runs 
for a given $f$. In Fig. \ref{Fig_11} the left panel shows the 
pdfs and cdfs of $b'$ for $f = 1$ for all the 10 runs. The pdf 
curves are akin and all have a slight decrease at $b \ge 0.95$. 
The mean of the pdfs is 0.0222 which is shown by the thick 
horizontal line. Lets assume a uniform distribution between [0, 1] 
with $k = 44$ bins then the mean of it is $1/k = 0.0227$ which 
is very close the observed value. Except from a random variation 
around the mean all the pdfs are similar to each other. Furthermore 
the cdf curves are essentially identical and has a slope of 1. 
These properties strongly indicates that the impact parameter has 
a uniform distribution within $[0, 0.95]$, for $b > 0.95$ the pdf 
drops off slightly. 

\par
As noted earlier the impact parameter has a notable influence on 
the collision outcome. For equal size bodies $\gamma = 1$ and 
$b_{\mathrm{crit}} = 0.5$ which is shown by a solid thick black 
vertical line on the left panel of Fig. \ref{Fig_11}, denoted by 
$\gamma_{1:1}$. According to the figure and make use of the result 
that $b$ has a uniform distribution the probability of a grazing 
impact is 
\begin{equation}
P(\mathrm{grazing}) = 1 - b_{\mathrm{crit}}.
\end{equation}
Assuming equal density the critical parameter can be express with $\gamma$ as
\begin{equation}
b_{\mathrm{crit}} = \frac{1}{1+\gamma^{1/3}},
\end{equation}
so the probability of a grazing impact as a function of $\gamma$ is
\begin{equation}
P(\mathrm{grazing}) = \frac{\gamma^{1/3}}{1+\gamma^{1/3}},
\label{Eq_P_grazing}
\end{equation}
which yields 50\% for equal size bodies. For 
e.g. $\gamma = 1:40$ the $b_{\mathrm{crit}} = 0.77$  
which is denoted by $\gamma_{1:40}$ on Fig. \ref{Fig_11}. 
This implies that less than 23\% of the collisions lead 
to grazing impact for bodies with a $\gamma \le 0.025$. 
As the protoplanetary disc matures and larger and
larger bodies emerge from the swarm $\gamma$ may reach
smaller values and the chance of a grazing impact 
reduces.

\par
The right panel of Fig. \ref{Fig_11} shows the pdfs and cdfs of 
$V_{\mathrm{i}}'$ for the same 10 runs. There are no collisions 
with impact speed less than 0.993, the maxima of the pdfs are at 
$\approx 1.07$, beyond it the pdfs drop off quite steeply at 
first, but then more slowly and finally the trend levels off 
around 5. Evidently there is a strong negative correlation, as 
$V_{\mathrm{i}}'$ gets larger than $\approx 1.07$ the relative 
frequency of collisions drop off quite steeply. From the figure 
it is obvious that the pdfs are very similar, while the cdfs 
look identical.

\subsubsection{Dependence on $f$}

\begin{figure}
\includegraphics[width=0.95\linewidth,keepaspectratio=true]{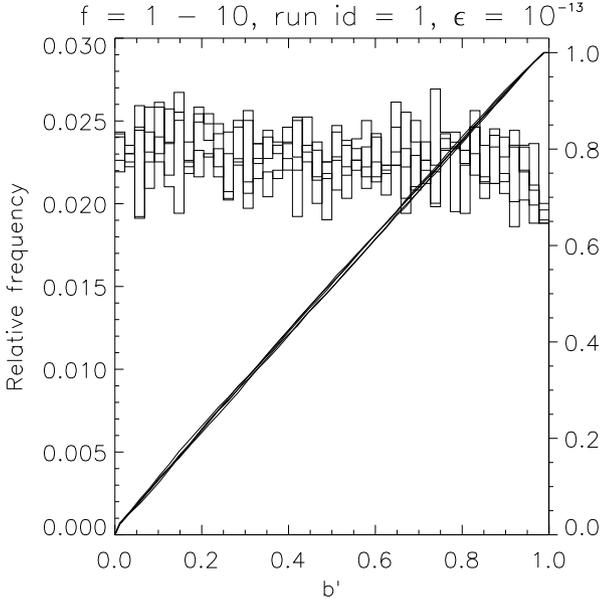}
\caption{The pdf and cdf of $b'$ for $\epsilon = 10^{-13}$ with 
run id = 1 for $f = 1,2,3,5,10$.}
\label{Fig_12}
\end{figure}

In this section the distribution of the collision parameters 
are compared for all the $f$ values. The distribution of the 
impact parameters is displayed in Fig. \ref{Fig_12} for 
$f = 1,2,3,5,10$ with $\epsilon = 10^{-13}$. Again, the pdf 
curves show a very similar behavior while the cdf curves 
are practically identical. It is evident that the impact 
parameter does not depend on $f$, it has a uniform distribution.

\par
\begin{figure}
\includegraphics[width=0.95\linewidth,keepaspectratio=true]{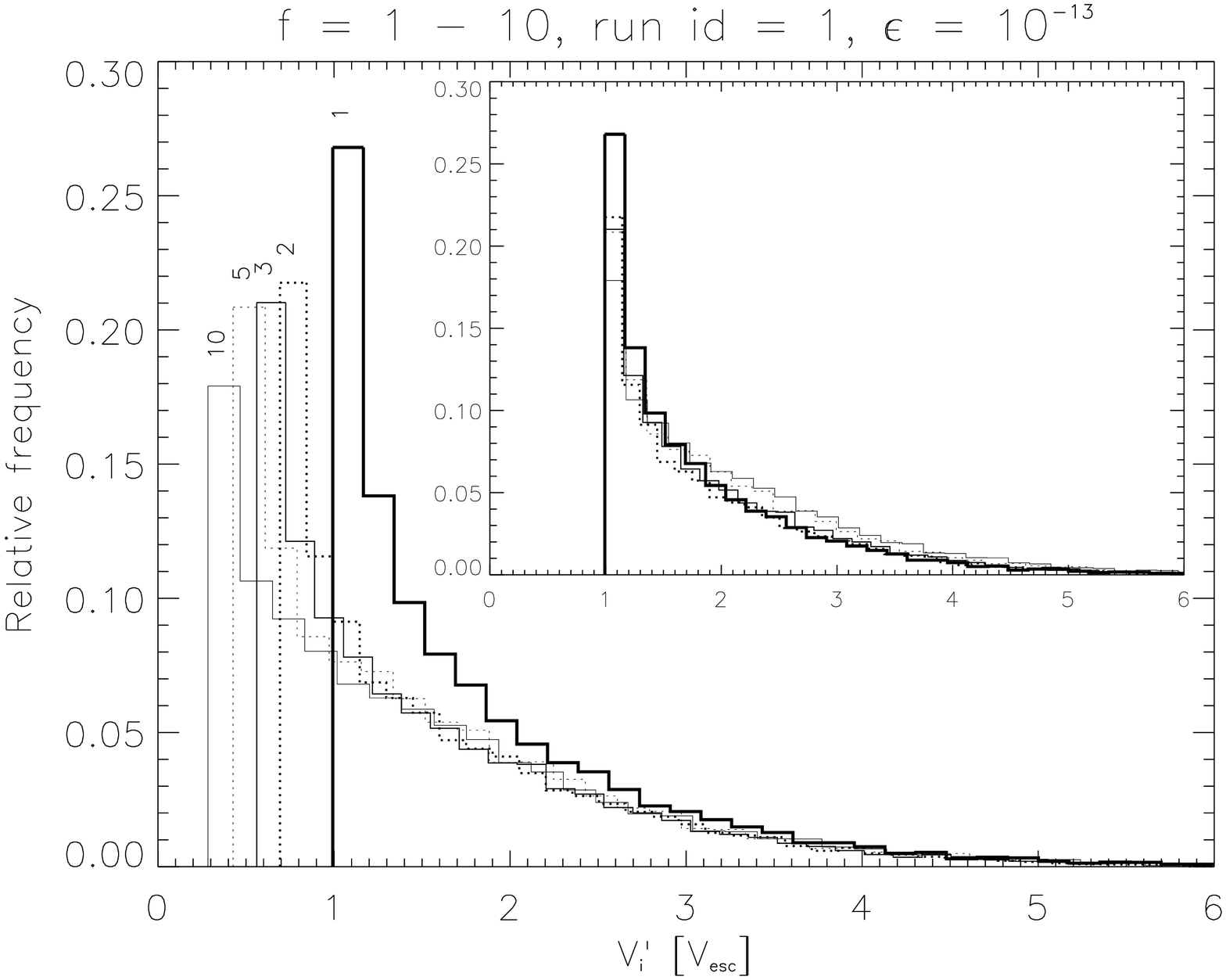}
\caption{The same as Fig. \ref{Fig_12} for $V_{\mathrm{i}}$. The thick 
solid black curve shows the pdf of $f = 1$. The inset plot shows the 
shifted pdfs; see text for details.}
\label{Fig_13}
\end{figure}

The distribution of the impact velocity for $f = 1,\ldots,10$ 
is shown in Fig. \ref{Fig_13} for $\epsilon = 10^{-13}$. 
Apparently the pdf of the impact velocity strongly depends on $f$. 
The domain of the pdf is equal 
to [min($V_{\mathrm{i}}$),\, max($V_{\mathrm{i}}$)] 
(see Table \ref{Tab_A1}) and min($V_{\mathrm{i}}$) shift towards 
lower values with increasing $f$, but the shape remains very 
similar as it is apparent from the inset plot, which was created
by shifting all the pdfs right to the location of minimum 
belonging to $f = 1$ which is approximately 1. The lower 
boundary of the domain, min($V_{\mathrm{i}}$) and location 
of the pdfs' maxima which is denoted by $V_{\mathrm{i,m}}$ 
are listed in Table \ref{Tab_09} and plotted in Fig. \ref{Fig_15} 
with asterisk and triangle, respectively. The smooth solid curve 
is derived from theoretical considerations detailed in the 
next section.

\begin{table}
\centering
\caption{For each $f$ the minimum value of the impact speed (\nth{2} column), 
the location of the pdfs' maxima (\nth{3} column), the calculated impact 
speed $V'_\mathrm{rel}$ from Eq. (\ref{Eq_v2p_approx}) 
(\nth{4} column) and the minimum distance $f_0$ and $d_0$ 
from Eq. (\ref{Eq_f0}) are listed for $\epsilon = 10^{-13}$.}
\begin{tabular}{rrrrrr}
\hline
 $f$ & min($V_{\mathrm{i}}$) & $V_{\mathrm{i,m}}$ & $V'_\mathrm{rel}$ & min($f_0$) & min($d_0$) [au] \\
\hline
 1 & 0.993 & 1.0801 & 1     & 71.68 & 2.40$\times 10^{-4}$ \\
 2 & 0.693 & 0.7686 & 0.707 & 50.63 & 1.69$\times 10^{-4}$ \\
 3 & 0.561 & 0.6435 & 0.577 & 53.73 & 1.80$\times 10^{-4}$ \\
 5 & 0.427 & 0.5181 & 0.447 & 56.59 & 1.89$\times 10^{-4}$ \\
10 & 0.284 & 0.3754 & 0.316 & 51.69 & 1.73$\times 10^{-4}$ \\
\hline
\end{tabular}
\label{Tab_09}
\end{table}

\section{Two-body approximation to model the impact velocity}

The observed behavior of the lower boundary of the domain and the 
location of pdfs' peak of the impact velocity can be explained by
a simple model described below.

\par
\begin{figure}
\includegraphics[width=0.95\linewidth,keepaspectratio=true]{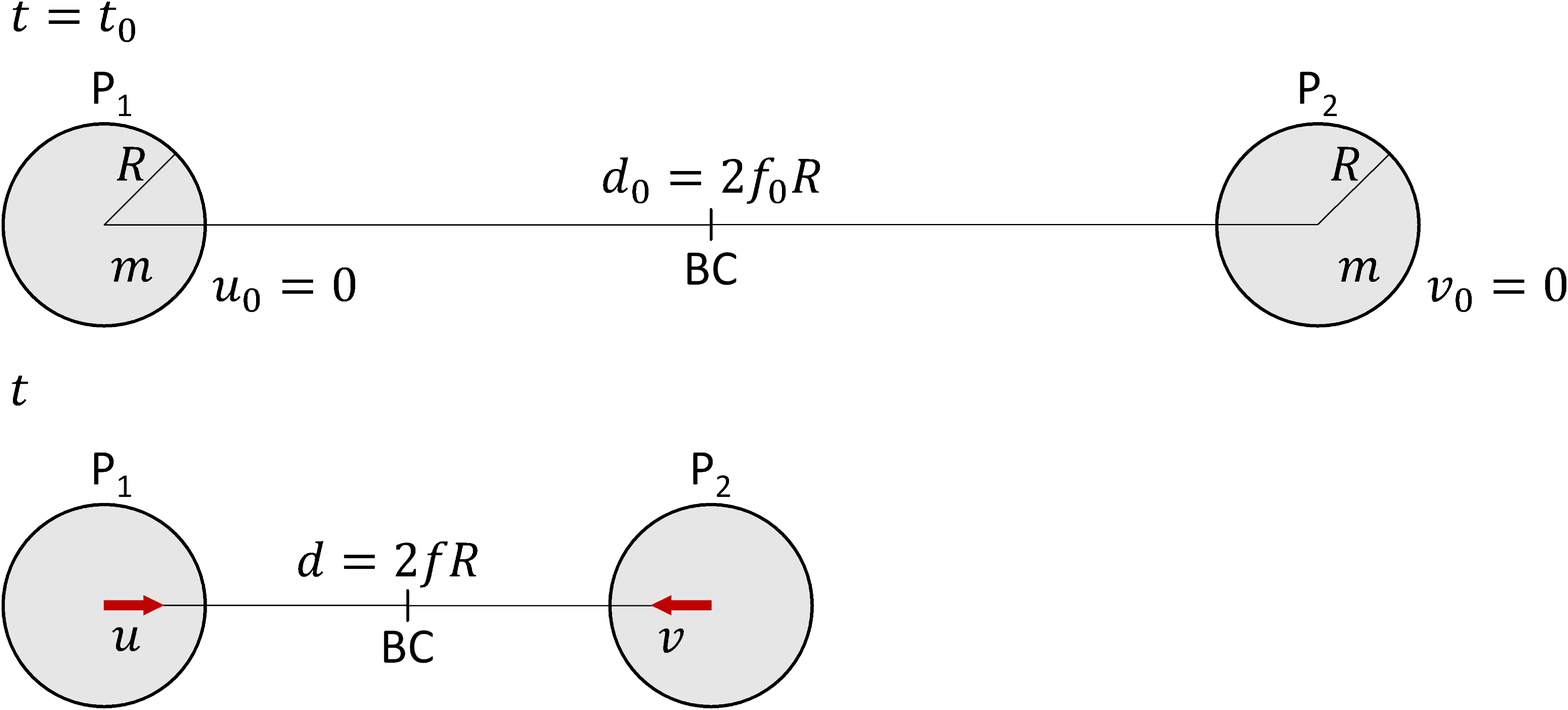}
\caption{The two-body problem in the center of mass reference frame.
BC denotes the barycenter.}
\label{Fig_14}
\end{figure}

I have applied a straightforward physical model based on the two-body
problem and used conservation laws to 
estimate the impact velocity as a function of the mutual distance $d$. 
In this model only the two colliding bodies $P_1$ and $P_2$ are considered 
and the effects of all others are neglected. In order to simplify the 
calculations I have assumed that the colliding bodies have the same 
mass $m$ and radius $R$. The model is depicted in Fig. \ref{Fig_14} 
where the coordinate system is fixed to the barycenter of the system 
denoted by BC in the figure.

\par
At time $t=t_0$ bodies $P_1$ and $P_2$ are $d_0$ apart from each other, 
and both bodies are in rest, $u_0 = v_0 = 0$. Let us calculate the 
velocity $u$ and $v$ of $P_1$ and $P_2$ at time $t$ when their distance 
is $d$. I note that the relative velocity plays the role of $V_{\mathrm{i}}$.
If $P_1$ is the target than the impact speed is $u-v$.

\par
From the conservation of the linear momentum it follows that $u=-v$ and from
the conservation of energy:
\begin{equation}
-k_\mathrm{G}^2\frac{m^2}{2f_0R} = \frac{1}{2}mu^2 + \frac{1}{2}mv^2 - k_\mathrm{G}^2\frac{m^2}{2fR},
\label{Eq_con_of_Energy}
\end{equation}
where I have parameterized the distance with $f \ge 1$ such that $d=2fR$.
Substituting $v = -u$ into Eq. (\ref{Eq_con_of_Energy}) one gets
\begin{equation}
u = k\sqrt{\frac{m}{2R}}\sqrt{\frac{f_0-f}{f_0f}}.
\label{Eq_u}
\end{equation}
The escape velocity of $P_2$ with respect to $P_1$ is 
\begin{equation}
V_\mathrm{e} = k\sqrt{2\frac{m}{R}},
\label{Eq_Vesc_BC}
\end{equation}
therefore $u$ can be written as
\begin{equation}
u = \frac{1}{2}V_\mathrm{e} \sqrt{\frac{f_0-f}{f_0f}}.
\label{Eq_u2}
\end{equation}
The relative velocity $V_\mathrm{rel}$ of $P_1$ with respect to $P_2$ is 
\begin{equation}
V_\mathrm{rel} = u - v = V_\mathrm{e} \sqrt{\frac{f_0-f}{f_0f}},
\label{Eq_V_rel}
\end{equation}
and introducing $V'_\mathrm{rel} = \frac{V_\mathrm{rel}}{V_\mathrm{e}}$
than Eq. (\ref{Eq_V_rel}) can be written as 
\begin{equation}
V'_\mathrm{rel} = \sqrt{\frac{f_0-f}{f_0f}}.
\label{Eq_v2p}
\end{equation}

\par
I remark that $V'_\mathrm{rel}$ is measured in escape velocity unit and
plays the role of the impact speed in the preceding text. 
If one knows the initial distance $d_0$ and thus $f_0$ than the 
velocity $V'_\mathrm{rel}$ at a distance of $d$ can be computed 
from Eq. (\ref{Eq_v2p}). If $f_0 \gg f$ then $f_0 - f \approx f_0$,
therefore
\begin{equation}
\frac{f_0 - f}{f_0f} \approx \frac{1}{f},
\label{Eq_approx}
\end{equation}
and then Eq. (\ref{Eq_v2p}) becomes
\begin{equation}
V'_\mathrm{rel} \approx f^{-1/2}.
\label{Eq_v2p_approx}
\end{equation}
This result gives an approximation of the minimum of the impact speed. 
Real physical impacts happen when $f = 1$, substituting this into 
Eq. (\ref{Eq_v2p_approx}) results $V'_\mathrm{rel} \approx 1$, i.e. 
the impact speed is approximately 1 escape velocity unit. This is 
exactly in line with the distribution of the impact velocities 
(see eg. Figs. \ref{Fig_08}, \ref{Fig_09}, and the right panel 
of Fig. \ref{Fig_11} and explains the lack of impact velocity less 
than about 1 for $f = 1$.

\par
\begin{figure}
\includegraphics[width=0.95\linewidth,keepaspectratio=true]{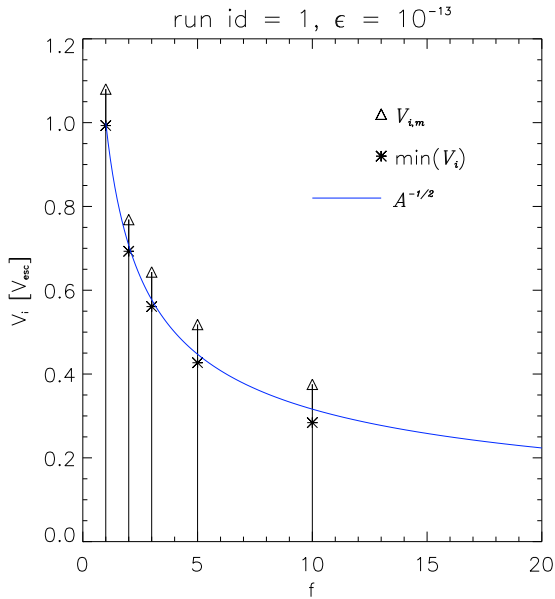}
\caption{The lower boundary of the domain min($V_{\mathrm{i}}$) and 
the location of the maximum $V_{\mathrm i, m}$ versus $f$ shown by 
asterisk and triangle symbols, respectively. The solid blue curve 
is the graph of the function in Eq. (\ref{Eq_v2p_approx}).}
\label{Fig_15}
\end{figure}

In Fig. \ref{Fig_15} the blue curve is the graph of Eq. 
(\ref{Eq_v2p_approx}). The min($V_{\mathrm{i}}$) data points match 
very well to this curve so this model explains nicely the shift 
observed on Fig. \ref{Fig_13}. From Fig. \ref{Fig_15} one can see 
that the minimum values are just below the blue curve, which is 
a natural consequence of that some of the bodies are close to 
each other, i.e. the assumption $f_0 \gg f$ breaks down. Substituting 
$f_0 > f = 1$ into Eq. (\ref{Eq_v2p}) one gets
\begin{equation}
V'_\mathrm{rel} = \sqrt{\frac{f_0 - 1}{f_0}} < 1.
\label{Eq_v2p_1}
\end{equation}
Using the formula of Eq. (\ref{Eq_v2p_approx}) for the different $f$ 
values the results are listed in the \nth{4} column of Table \ref{Tab_09}.

\par
The initial distance characterized by $f_0$ can be expressed from 
Eq. (\ref{Eq_v2p})
\begin{equation}
f_0 = \frac{f}{1 - f V'^2_\mathrm{rel}}.
\label{Eq_f0}
\end{equation}
Substituting the minimum of the impact speed into $V'_\mathrm{rel}$ from 
the \nth{2} column of Table \ref{Tab_09} into Eq. (\ref{Eq_f0}) 
then the initial distance of the two bodies can be estimated and 
the resulting $f_0$ is displayed in the \nth{5} column of Table 
\ref{Tab_09}.

\section{Collision outcome maps}
\label{section_collision_outcome_maps}

\begin{figure*}
\includegraphics[width=0.48\linewidth,keepaspectratio=true]{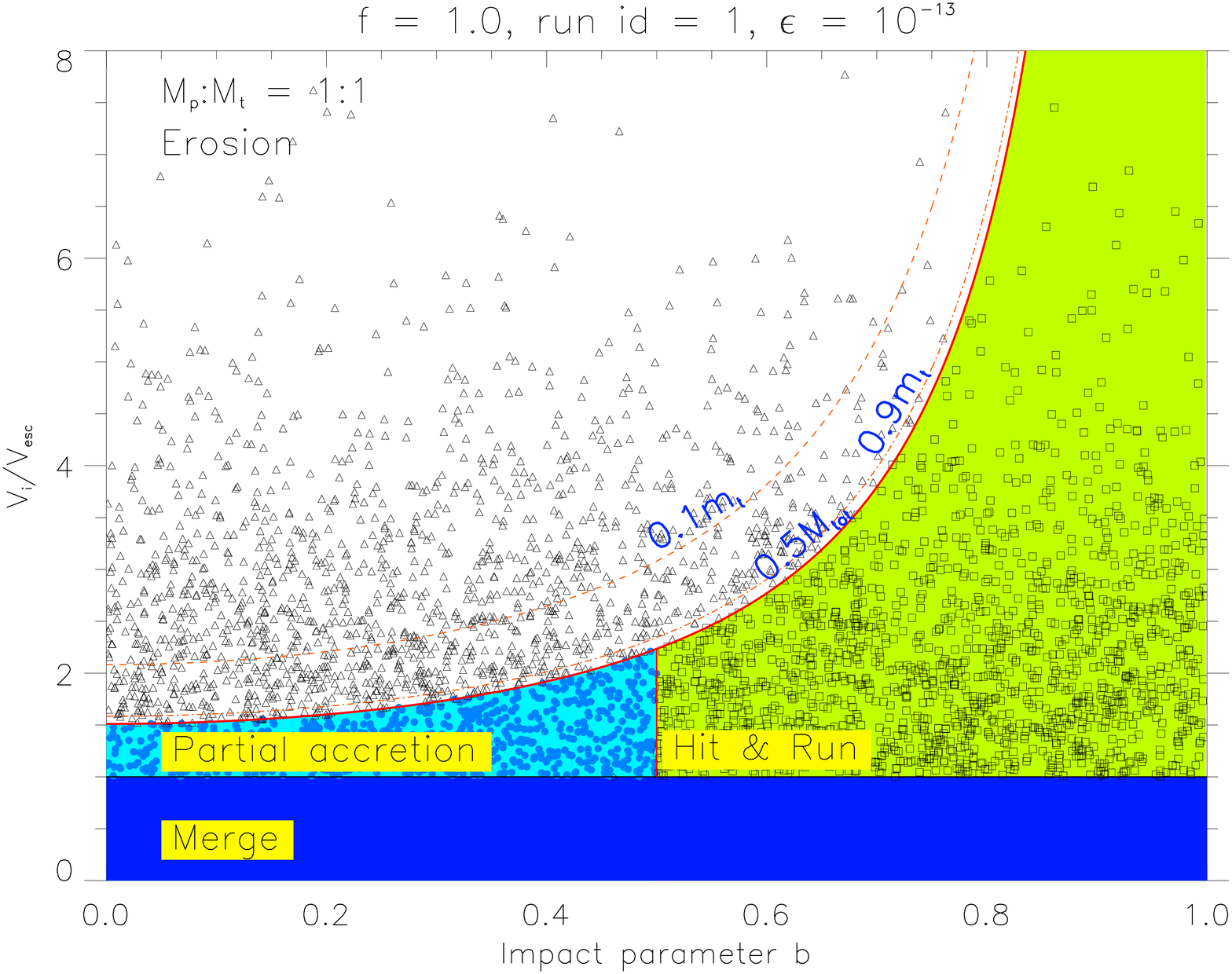}
\includegraphics[width=0.48\linewidth,keepaspectratio=true]{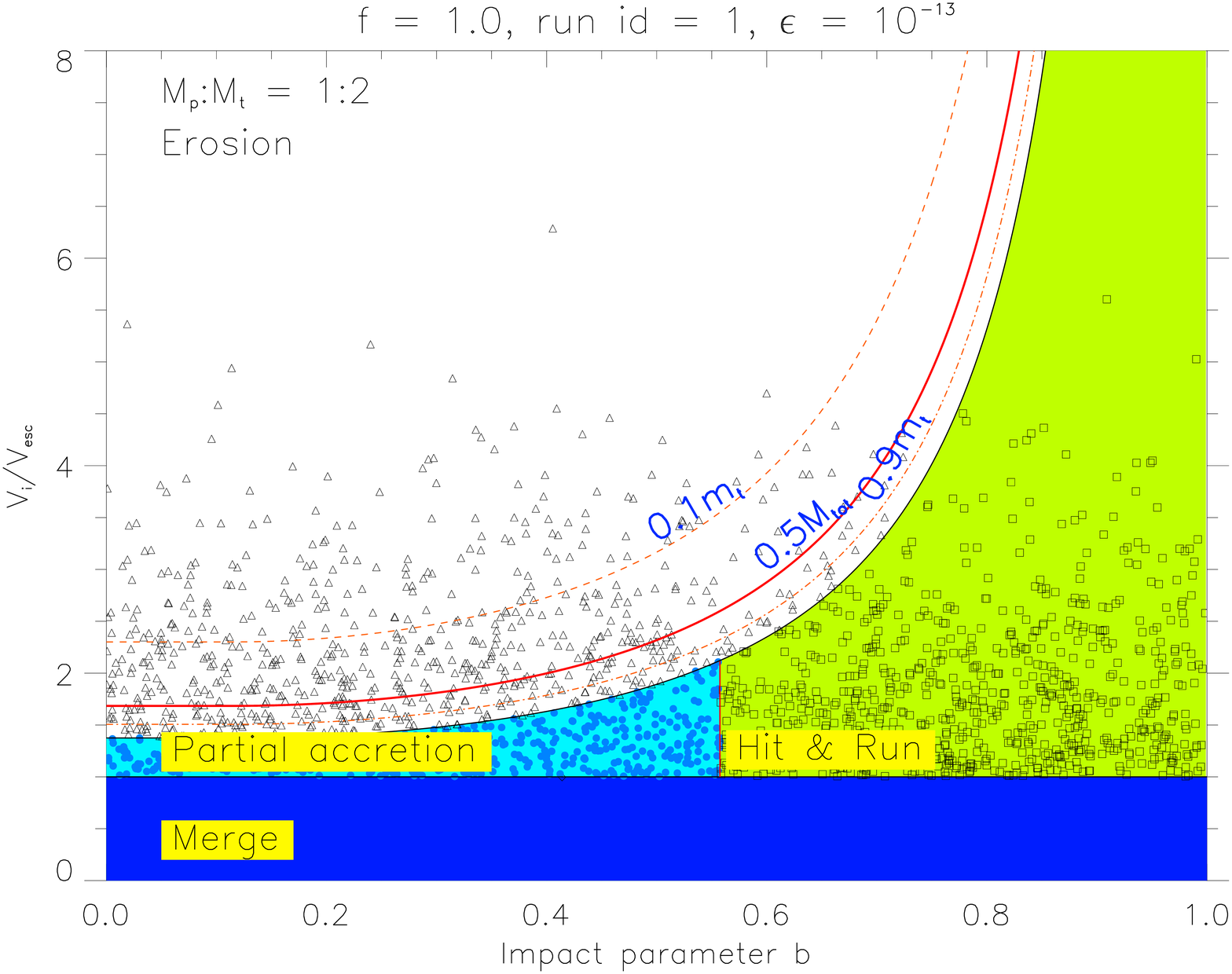}
\caption{Predicted collision outcome maps using the 
analytic model of LS12 for strengthless planets ($
\bar{\mu}= 0.36$ and $c^* = 1.9$). Impact velocity
is normalized by the mutual surface escape velocity
assuming a bulk density of 2000 kg m$^{-3}$. Colored 
regions denote perfect merging (dark blue), 
partial accretion (light blue), net erosion to the target 
(white), and hit-and-run (green). The vertical red line 
denotes the onset of hit-and-run events at 
$b_{\mathrm{crit}}$. Thick black red denotes the 
critical disruption velocity for half the total mass 
($0.5M_{\mathrm{tot}}$) remaining; red dashed curve 
denotes 90\% ($0.9m_t$) and 10\% ($0.1m_t$) of 
target mass in largest remnant. The different symbols 
indicate the outcome of each collision: diamonds for 
mergers, squares for hit-and-run collisions, 
filled circles for collisions in which the target gained 
mass, and triangles for collisions in which the target 
lost mass.}
\label{Fig_16}
\end{figure*}

Recent works (LS12, \cite{Genda2012}) based on the combination of 
hydrocode and $N$-body gravity code computations studied the impacts 
between planetary mass bodies and the outcome of planetary collisions   
have been parameterized  in terms of the masses and velocities of the
colliding bodies. According to these calculations, the authors devised 
formulae for the mass of the largest remnant denoted by 
$M_{\mathrm{lr}}$ produced in a collision as a function 
of the masses of the bodies involved, the impact velocity 
and impact angle. The authors also identified the 
boundaries between different types of collision:
\begin{enumerate}
\item perfect merging ($M_{\mathrm{lr}} = 
M_{\mathrm{tot}}$) 
\item partial accretion with some mass escaping as 
fragments ($M_{\mathrm{lr}} < M_{\mathrm{tot}}$)
\item partial erosion of the target ($M_{\mathrm{lr}} < 
m_{\mathrm{t}}$)
\item pure hit-and-run ($M_{\mathrm{lr}} = m_{\mathrm{t}}
$) and erosive hit-and-run (lead to some erosion of the 
target and more significant damage of the projectile.)
\end{enumerate}
where $M_{\mathrm{tot}} = m_{\mathrm{t}} + m_{\mathrm{p}}$.

\par
Using the analytic model of LS12 with adopted values of 
$\bar{\mu}= 0.36$ and $c^* = 1.9$, I derived example
collision outcome maps. I present an analysis of the 
simulations to assess the range of true collision outcomes. 
For this purpose two color-coded collision outcome maps 
were calculated which are shown in Fig. \ref{Fig_16} for 
mass ratios of $\gamma = 1$ (left panel) and $\gamma = 1:2$ 
(right panel). In the figure dark blue denotes perfect merging,
light blue partial accretion, white color net erosion to the 
target and finally green hit-and-run region. In every case, 
the simulation data were used to calculate the collision 
parameters and to determine the outcome based on the LS12 model.
The vertical red line denotes the onset of hit-and-run 
events at $b_{\mathrm{crit}}$.

\par
The details of the calculation of the collision regimes 
are given in the Appendix of LS12. The model assumes a 
sudden transition between grazing and non-grazing 
impacts, which is of course, artificial. In Fig. 
\ref{Fig_16}, the thick red curve corresponds to the 
critical velocity for catastrophic disruption, where the 
largest remnant contains half the total mass, 
$M_{\mathrm{lr}} = 0.5M_{\mathrm{tot}}$. Note that this 
curve corresponds to the target erosion boundary for $
\gamma = 1$ cases (the transition from partial accretion 
or hit-and-run to the erosion region). The lower red 
dot-dashed curves correspond to the impact velocity 
needed to disperse 10\% and the upper red dashed curve 
to 90\% of $m_{\mathrm{t}}$.

\par
Fig. \ref{Fig_16} also shows details of all the 
collisions that occurred in the simulations for $f = 1$ 
and $\epsilon = 10^{-13}$ for a single run data set, run 
id = 1 with each symbol representing a single collision 
with $\gamma = 1$ (left) and $\gamma = \frac{1}{2}$ (right).
The symbols indicate the type of collision involved: 
diamonds for mergers, squares for hit-and-run collisions 
where the target survives intact, circles for collisions 
that increased the mass of the target body (possibly with 
some mass from the projectile escaping as fragments), and 
triangles for collisions that eroded mass from the 
target. From the figure it is apparent that there is no 
correlation between $V_{\mathrm{i}}$ and $b$. The number 
of the different type collisions are summarized 
Table \ref{Tab_10}.

\par
In Table \ref{Tab_10} the total number of collisions $n$ and 
the ratios of the different types of outcomes are given 
for five $\gamma$ values. The number of perfect merging, 
partial accretion, erosion and hit-and-run events are denoted by 
$n_{\mathrm{m}}$, $n_{\mathrm{pa}}$, $n_{\mathrm{e}}$ and
$n_{\mathrm{hr}}$, respectively. In the table 
the subscript e1 denotes collisions where 
$M_{\mathrm{lr}} \le 0.5M_{\mathrm{tot}}$, i.e. the 
catastrophic disruption, e2 denotes 
collisions where $M_{\mathrm{lr}} \le 0.9m_{\mathrm{t}}$ 
and e3 denotes collisions where $M_{\mathrm{lr}} \le 
0.1m_{\mathrm{t}}$, i.e. the super-catastrophic collision. 
The number of merging collisions is very low, less than 1\% 
for all listed $\gamma$ values.

\begin{table}
\centering
\caption{Predicted collision outcome ratios in percentages for
different $\gamma$ with $f = 1$ and $\epsilon = 10^{-13}$. The total 
number of listed cases is 67716 representing 68\% of the sample.}
\begin{tabular}{rrrrrr}
\hline
  & 
  $\gamma = 1$ & $\gamma = \frac{1}{2}$ & $\gamma = \frac{1}{3}$ & $\gamma = \frac{1}{4}$ & $\gamma = \frac{1}{10}$ \\
\hline
$n$                 & 36656 & 18564 & 7866  &  4088 &   542 \\
$n_{\mathrm{m}}/n$  &  0.05 &  0.11 &  0.22 &  0.24 &  0.55 \\
$n_{\mathrm{pa}}/n$ & 17.16 & 24.21 & 32.43 & 37.77 & 63.47 \\
$n_{\mathrm{e}}/n$  & 39.79 & 35.35 & 27.77 & 24.00 &  6.27 \\
$n_{\mathrm{hr}}/n$ & 43.00 & 40.33 & 39.59 & 37.99 & 29.70 \\
\hline
$n_{\mathrm{e1}}/n$ & 39.79 & 23.43 & 10.39 &  5.04 &  0.37 \\
$n_{\mathrm{e2}}/n$ & 37.39 & 29.83 & 20.21 & 14.36 &  0.92 \\
$n_{\mathrm{e3}}/n$ & 23.29 &  9.87 &  2.81 &  0.76 &  0.00 \\
\hline
\end{tabular}
\label{Tab_10}
\end{table}

\par
\begin{figure}
\includegraphics[width=0.95\linewidth,keepaspectratio=true]{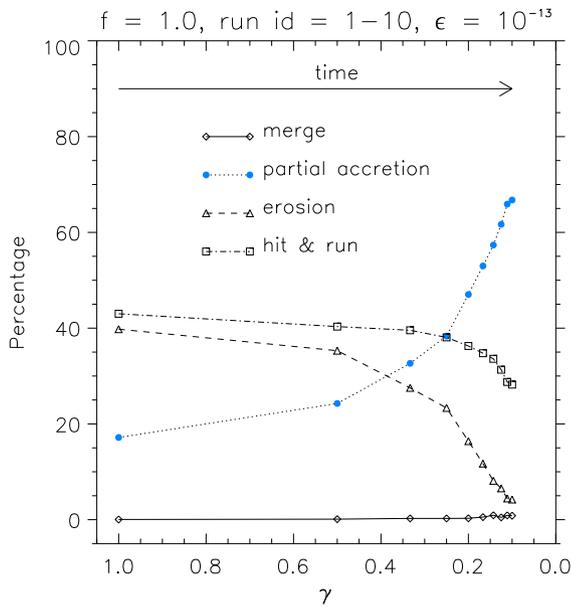}
\caption{The frequency of different type of collision as a function of 
$\gamma$. Note that the horizontal axis is reversed in order to reflect 
the progress of time, as it is shown by the upper arrow. The symbols 
indicate the type of collision: diamonds for mergers, blue circles for 
partial accretion, triangles for collisions that eroded mass from the 
target, and squares for hit-and-run. The collisions were collected 
from a multiple run data with $f = 1$ and $\epsilon = 10^{-13}$.}
\label{Fig_17}
\end{figure}

According to the table there is a clear trend which is 
visualized in Fig. \ref{Fig_17}, where  I have plotted 
the number of the different type collisions against
$\gamma$. 

The initial single mass distribution relaxes into
a continuous power-law mass distribution in ${\sim}100$
years which is in line with the results of 
\cite{Kokubo1996}. As the simulation time proceeds the
largest body of the continuous mass distribution 
separates from it. In the beginning of the simulation 
$\gamma = 1$ and as the continuous power-law mass 
distribution develops the decreasing minimum of $\gamma$ 
reflects the time evolution of the system, at
least early in the simulation. Later on this relationship 
breaks and all we can say is that the minimum value
of $\gamma$ is indicative of the minimum age of the 
system. In order to reflect this initial relationship 
between time and min($\gamma$) the horizontal axis $\gamma$ 
is reversed and the time is shown by the black arrow in 
the top of the figure.

\par
It is clearly shown in Fig. \ref{Fig_17} that the 
number of merging collisions remains very low, below 1\% 
and shows a very modest increase as $\gamma$ decreases. 
As time proceeds and the bodies grow via collisions 
$\gamma$ may reach smaller values therefore in later 
times the chance of a merging collision slightly increases.
The number of partial accretion depends strongly on 
$\gamma$. As min($\gamma$) decreases the chance of 
partial accretion events increases 
and for $\gamma \le 0.5$ the increase speeds up such that 
the proportion of partial accretion reaches $\approx 70\%$ 
for $\gamma = 0.1$. On the other hand the number of 
erosion declines with decreasing $\gamma$ just as 
$n_{\mathrm{hr}}$ which follows a similar pattern but it 
decreases with a smaller pace. If the collision rate
is maintained by gravitational focusing and/or additional 
material is coming from outer space via migration than 
$\gamma$ may reach smaller values and the chance of a 
grazing impact reduces, see Eq. (\ref{Eq_P_grazing}). 
For $\gamma < 0.2$ partial accretion dominates and 
consequently as the disc matures and time increases the 
larger bodies grow faster, i.e. the rich get richer.

\par
The importance of hit-and-run collisions was recognized 
by \cite{Asphaug2006}, namely in these events, a pair of 
objects collides at an oblique angle. Momentum causes the 
bodies to graze each other while exchanging some mantel 
material and afterwards they separate again. Generally, 
the final bodies have similar masses to the original pair 
but smaller relative velocities. Hit-and-run collisions 
can be further divided into two subclasses. When the 
first collision speed is only slightly larger than the 
mutual escape velocity, the bodies collide, separate but 
in a second  collision they merge. These are the so called 
graze-and-merge collisions. On the other hand when the 
initial impact velocity is higher, the two bodies separate 
with a relative speed larger than the escape velocity, 
which is not followed by a second impact. These are true 
hit-and-run collisions. The boundary between these two 
regimes has been clearly identified in a second series of 
impact simulations carried out by \cite{Genda2012}.

\par

According to the definition of \cite{Kokubo1996} runaway 
growth means that the largest body in its feeding zone 
grows more rapidly than the second largest one and the 
ratio of the mass of the largest body $M_{\mathrm{max}}$ 
and the mean mass $\langle m \rangle$ grows monotonically 
with time. In their 3D calculations they presented 
evidence of runaway growth and the results in this work 
summarized in Fig. \ref{Fig_17} also supports this 
observation. As the single mass distribution relaxes into
the continuous power-law mass distribution more and more 
collisions may involve a larger and smaller body, i.e. $
\gamma \le 0.2$ hence partial accretion dominates 
therefore the larger the body is the faster it grows.
The numerical simulations of \cite{Aarseth1993} showed 
that runaway coagulation occurs and can lead to the 
formation of protoplanetary cores. In their 2D 
simulations in the case of larger eccentricities (hot 
accretion) the degree of runaway growth can be the same 
as in 3D. In the 3D simulations of \cite{Kokubo1996} the
ratio $M_{\mathrm{max}} / \langle m \rangle \approx 140$ at 
$t = 20 000$ yr, in their 2D one the ratio 
$\approx 15$, while in the work of \cite{Aarseth1993} it 
was $\approx 13$. Using the data of the present work 
preliminary calculations show that 
$M_{\mathrm{max}} / \langle m \rangle \approx 10$ at $t = 
20 000$ yr.
It must be noted that the above mentioned works can not 
be directly compared because on one hand of the 
significant differences between the treatment of the
gravitational forces and the collisions and on the other 
hand the different initial conditions. However, the 
general conclusions and trends are similar.
In the present work the main focus is on the statistics
of the collision parameters and hence growth is studied 
from the collision point of view, therefore these results 
are valid for both 2D and 3D cases. The issue about the
orderly and runaway growth as well as the mass 
distribution of the bodies and its time evolution will be 
investigated in a next paper.

\subsection{Comparison with previous works}

A direct comparison to previous works is hard to present because of the
different simulation setups. The most important is that the present 
simulations were performed in 2D while the others in 3D. The
initial conditions, the $f$ expansion factor of the radii differs
from work to work and the values in the tables or figures are given 
for different $\gamma$ or $\gamma$ is not exactly specified in 
the papers.

\par

To estimate the effects of including the collision model of LS12 on
planet formation \cite{Stewart2012} analyzed the impact parameters 
of previous $N$-body simulations that used only perfect
accretion \citep{Obrien2006,Raymond2009}. The authors
divided the collision outcomes from the simulations into 3 groups: 
group 1 consists of all collisions from eight different simulations in 
\cite{Obrien2006}, group 2 includes all collisions from 40
simulations by \cite{Raymond2009} while group 3 
contains only giant impacts from the latter simulations. Group 1 and 2 
were subdivided into two subclass containing collisions between 
embryos and planetesimals ($\gamma \le \frac{1}{40}$) and between 
only embryos ($\gamma \ge \frac{1}{10}$). The predicted collision
outcome statistics from Table 1 of \cite{Stewart2012} is summarized in the columns
from 5 to 9 in Table \ref{Tab_11}.

\par
The simulations of \cite{Chambers2013} also include the LS12
model to take care of fragmentation and hit-and-run collisions.
The collision statistics of these calculations using Table 1 and 
Figs. 5 and 6 of \cite{Chambers2013} are reported in the \nth{10} and
\nth{11} columns of Table \ref{Tab_11}.

\par
In a recent work \cite{Bonsor2015} performed $N$-body simulations to track 
the change in the bulk Mg/Fe and Si/Fe ratios of the protoplanets. The
$N$-body code used the LS12 model to realistically model collisions. However,
the paper did not present the collision statistics in a tabular form, this
kind of data could be extracted from Figure 4 of the paper. The approximate
values are presented in the last 2 columns of Table \ref{Tab_11}. Similar
runs were performed by \cite{Carter2015} and the collision statistics 
derived from Figure 5 of the paper is qualitatively the same as in 
\cite{Bonsor2015}.

\par
The impacts for multiple run data with $f=1$ and $\epsilon = 10^{-13}$
of this work is presented in the first three columns of Table \ref{Tab_11},
where the \nth{2} column list data for all $\gamma$, \nth{3} column
for giant impacts with $\gamma \ge \frac{1}{10}$ and the \nth{3} column 
shows for planetesimal impacts with $\gamma \le \frac{1}{40}$. 
These specific values were selected since the initial $\gamma$ 
between the planetesimals and embryos were similar in 
\cite{Stewart2012}.

\par
Comparing the giant impacts of this work (\nth{3} column of 
Table \ref{Tab_11})
with that of \cite{Stewart2012} (\nth{5} -- \nth{7} columns) one see similarities
and significant differences. The perfect merge ratios roughly agree in all 
cases except in group 1 where 0 merge event was detected, but the sample size
in that group is at least an order of magnitude smaller than in the other 
ones. The hit-and-run case represents 29.9\% of all events in the
present work and this value agrees well with that of group 1 (31.3\%),
2 (27.7\%) and 3 (28.2\%). The partial accretion measured in this work is 
typically larger by a factor of $\approx 1.5$, the best agreement is with 
group 2: 44.7\% vs 39.2 \%. Significant differences were found 
in the case of erosion: the present study found 24.9\% while 
this quantity is less than 3\% in all the groups. The reason for
this large difference is unclear since the impact velocity distribution
measured in this study (see Figs. \ref{Fig_10}, \ref{Fig_11},
\ref{Fig_12} and \ref{Fig_15}) 
and those presented in \cite{Stewart2012} (Figs. 2 and 10) show good agreement.
This large discrepancy deserves careful further studies since erosion plays
a crucial role in the formation and the final bulk composition of the
terrestrial planets. Notwithstanding, the special case of erosion the
super-catastrophic events ratio of the present work (0.2\%) matches 
quite well with the value in group 3 (0.3\%).

\par
Analyzing the planetesimal collisions detected in the present work 
(see the \nth{4} column) and that of \cite{Stewart2012} 
(\nth{8}, \nth{9} columns) a very good agreement can be observed: 
partial accretion 76.8\% vs. 72.6\% and 69.4\%, hit-and-run events 22.2\%
vs 23.6\% and 25.4\% and there was no super-catastrophic event
in all three simulations.

\par
Comparing all impacts of this work (\nth{2} column) with that of 
\cite{Chambers2013} (\nth{10} column) one sees that in  
overall the probabilities of different collision outcomes are similar.
The results for partial accretion and erosion should be highlighted, 
29.8\% vs 32.4\% and 24.1\% vs 19.7\%, respectively. These present
a far better agreement then in the previous comparison. Also the
hit-and-run events match closely: 45.7\% vs 44.3\%. On the other hand
the comparison of giant impacts (\nth{3} column vs \nth{11}) shows 
significant differences: the perfect accretion calculated in the
present work is 0.3\% much less than the 25\% reported by 
\cite{Chambers2013}, the erosion in this work is 24.9\% vs 0\%. The
values for partial accretion and hit-and-run broadly agrees. It
must be noted that in the work of \cite{Chambers2013} $\gamma$ was
not specified, these values were derived from the embryo-embryo 
collisions taken from the last row of Table 1 in \cite{Chambers2013}.
Since the collision outcome depends strongly on $\gamma$ 
these values are only approximate and therefore they are only
indicative.

\par
The last 2 columns contains the collision statistics compiled from 
Figs. 4 and 5 of \cite{Bonsor2015} both containing collision events
with $\gamma \in (0,1]$. The $t=0$ means the beginning of the simulation
while $t = 6\times 10^5$ is the end of it. These values should be
compared with the \nth{2} column and one see that in the present work 
perfect merging is about 20 times less than in \cite{Bonsor2015}. 
The partial accretion is almost equal, while the hit-and-run ratio
of this work is 45.7\% which is in between the values of 50\% and 36\% 
at $t = 0$ and $t = 6\times 10^5$, respectively. Again the erosion
is largely different: at $t = 0$ it is 24\% vs 3\%, but this difference
decreases and the erosion reaches 16\% at the end of the simulation.

\par
Comparing the works enumerated in Table \ref{Tab_11} with each other 
the differences are large and the results vary in large intervals. 
On one hand this is a natural consequence of the different simulation 
setup, applied numerical integrator method and expansion factor but 
on the other hand this 
issue has to study further in detail; especially the cause behind
the large difference in the measured erosion and perfect accretion.

\par
I note that the model of LS12 applies a different definition for the escape velocity:
\begin{equation}
V'_\mathrm{esc} = k_\mathrm{G}\sqrt{2M'/R'},
\end{equation}
where $M' = m_\mathrm{t} + m_\mathrm{interact}$. The $m_\mathrm{interact}$ denotes 
the interacting mass of the of the projectile estimated to be involved in the collision.
Since $m_\mathrm{interact} \le m_\mathrm{p}$ therefore it may be a contributing factor in 
the difference of perfect mergers in \cite{Bonsor2015} and \cite{Chambers2013}.

\begin{table*}
\centering
\caption{Predicted collision outcome ratios calculated from the work
of \protect\cite{Stewart2012,Chambers2013,Bonsor2015} and from the present
simulations. The number of graze-and-merge events is denoted by $n_{\mathrm{gm}}$.}
\begin{tabular}{r|rrr|rrrrr|rr|rr}
\hline 
 & \multicolumn{3}{c|}{This work} & \multicolumn{5}{c|}{\protect\cite{Stewart2012}} & \multicolumn{2}{c|}{\protect\cite{Chambers2013}} & \multicolumn{2}{c}{\protect\cite{Bonsor2015}} \\
 & \multicolumn{3}{c|}{} & \multicolumn{5}{c|}{[Table 1]} & \multicolumn{2}{c|}{[Table 1, Figs. 5 \& 6]} & \multicolumn{2}{c}{[Figs. 3 \& 4]} \\
 & All & Giant & Planet. & \multicolumn{3}{c}{Giant} & \multicolumn{2}{c|}{Planetesimal} & all coll. & all coll. & $t=0$ & $t=6\times 10^5$ \\
 & $\gamma \in (0,1]$ & $\gamma \ge \frac{1}{10}$ & $\gamma \le \frac{1}{40}$ & \multicolumn{3}{c}{$\gamma \ge \frac{1}{10}$} & \multicolumn{2}{c|}{$\gamma \le \frac{1}{40}$} & (1 sim) & (em - em) & all coll. & all coll. \\
\hline
$n$                 & 99899 & 93330 &  1530 &    67 &   544 & 1165 &  1140 &  3142 & $\approx 386$ &   92 &  -- &  -- \\
$n_{\mathrm{m}}/n$  & 0.32  &   0.3 &   0.9 &   0.0 &   0.7 &  0.6 &   0.0 &   0.6 &           3.6 &   25 &   6 &   5 \\
$n_{\mathrm{pa}}/n$ & 29.82 &  44.7 &  76.8 &  26.9 &  39.2 & 36.1 &  72.6 &  69.4 &          32.4 &   37 &  41 &  43 \\
$n_{\mathrm{e}}/n$  & 24.14 &  24.9 &   0.1 &   3.0 &   0.6 &  1.3 &   0.0 &   1.9 &          19.7 &    0 &   3 &  16 \\
$n_{\mathrm{hr}}/n$ & 45.72 &  29.9 &  22.2 &  31.3 &  27.7 & 28.2 &  23.6 &  25.4 &          44.3 &   38 &  50 &  36 \\
$n_{\mathrm{gm}}/n$ & 	--  &    -- &    -- &  38.8	&  31.8	& 33.8 &   3.8 &   2.7 &	        -- &   -- &	 -- &  -- \\
$n_{\mathrm{e3}}/n$ & 12.07 &   0.2 &   0   &   1.5 &   0   &  0.3 &   0.0 &   0.0 &            -- &   -- &   1 &  10 \\
\hline
Notes. & \multicolumn{12}{l}{In the work of \protect\cite{Chambers2013} the $n_{\mathrm{m}}/n$ contains also the graze-and-merge events.} \\
\end{tabular}
\label{Tab_11}
\end{table*}

\section{Conclusions}

I have performed 2D $N$-body simulations of planetary accretion. 
The statistics of the collision parameters is investigated with 
10000 equal-mass protoplanets under perfect and gas-free accretion. 
In this paper the detailed statistics of the collision parameters 
are presented along with a simple method to improve the collision 
parameters. Using the two-body problem and the conservation laws
I explained the main features of the observed impact velocity 
distributions. The collision outcome maps are shown 
for specific projectile-to-target mass ratios. Combining the 
results of the simulations with the model developed by LS12 
estimates for the different type of collisions are given as well as
detailed comparisons with previous works on collision outcomes
and frequencies.

\par
The main conclusions of this study are:
\begin{itemize}
\item Using the two body approximation for the colliding bodies 
a simple method was presented to improve the impact parameter 
$b$ to $b'$ by Eq. (\ref{Eq_bprime}) and the impact velocity 
$V_{\mathrm{i}}$ to $V_{\mathrm{i}}'$ by Eq. (\ref{Eq_Vi_prime}). 
Using $V_{\mathrm{i}}'$ the improved specific energy $Q_{\mathrm{R}}'$ 
can be computed.

\item It is apparent from Figs. \ref{Fig_08} and \ref{Fig_09} panel b) 
that the pdf curves of $b$ and $b'$ are similar and from a statistical 
point of view they do not deviate, i.e. the improvement of the impact 
parameter is not too remarkable in their pdfs. Similarly the pdfs of 
the improved impact velocity from panel c) of the same figures show 
little difference, although the largest discrepancy is for the most 
important interval. These findings are also supported by Fig. \ref{Fig_10}, 
where the results are directly compared for $\epsilon = 10^{-10}$ 
and $\epsilon = 10^{-13}$. I emphasize that the correction is important 
for the individual cases and it is obvious that the improvement 
of $b$ and $V_{\mathrm{i}}$ is essential and all future simulations 
that use the model of LS12 (or similar) should incorporate it to 
provide more reliable results.

\item It was shown that in 2D the impact parameter has a uniform 
distribution within $[0, 0.95]$, and for $b > 0.95$ the pdf drops 
off slightly.

\item According to Fig. \ref{Fig_11} the different runs produce the 
same distribution of $b'$, $V_{\mathrm{i}}'$ and $Q_{\mathrm{R}}'$
therefore it is enough to do only one run to obtain creditable 
statistical data.

\item Making use of the two body approximation it was possible 
to explain the hiatus of impact speed less than 1 $V_{\mathrm{esc}}$ for
$f=1$ which is a consequence of the conservation laws. The
derived formula gives an excellent approximation of the minimum 
impact speed for different $f$.

\item It was shown that the impact parameter $b$ does not depend on $f$ 
as it is presented by Fig. \ref{Fig_12}. On the other hand the 
distribution of $V_{\mathrm{i}}$ is a strong function of $f$, the 
pdf of the impact speed shifts toward zero as $f$ increases, see 
Fig. \ref{Fig_13}. This behaviour was explained by the two 
body approximation shown in Fig. \ref{Fig_15}.

\item The model of LS12 was used to determine the number of different 
types of collisions, see Fig. \ref{Fig_16} and Table \ref{Tab_10}. 
For equal-mass bodies the majority of collisions (43\%) are hit-and-run 
events and erosion (40\%). Partial accretion constitutes 17\% of 
all cases and 0.05\% mergers were measured. The frequencies of
different events are summarized in Fig \ref{Fig_17}.

\item A thorough comparison 
with previous works was presented and promising similarities along with
significant differences were found. The most significant 
difference is in the case of erosion, where this work 
reports 25\%, while all the others are much smaller, less than 3\%
expect for one case of \cite{Chambers2013} where it reaches approximately 20\%.
The ratio of perfect merge roughly agree with those of \cite{Stewart2012}
but differs largly from the results of \cite{Chambers2013,Bonsor2015}.
The hit-and-run values match reasonably, while the partial accretion ratios
broadly agree with other values. It was also shown that serious differencies 
are present among the previous works.

\item The proportion of partial accretion increasing more 
and more steeply as min($\gamma$) decreases, i.e. as the 
time increases, see Fig. \ref{Fig_17}. For $\gamma \le 0.1$ the 
majority of collisions ($\approx 70$\%) is partial accretion 
and 25\% is hit-and-run events (comparable values were found
by \cite{Stewart2012}). A significant fraction of the 
latter events are graze-and-merge collisions which also leads to 
accretion. These results provide a independent evidence for the 
runaway growth mode described in detail by \cite{Kokubo1996}. The 
implications for planet formation is that the larger the difference 
between the masses of the objects ($\gamma$ is small) the larger 
the probability for collisions in which $M_{\mathrm{lr}}$ grows. 
This can further decrease min($\gamma$) and thus induce a positive 
feedback favoring more intensive mass growth.
\end{itemize}

\par
Below I summarize some technically important findings:
\begin{itemize}
\item The run-times for the two accuracy parameters were compared 
and it was shown that the simulations for $\epsilon = 10^{-13}$ takes 
about 40\% more time to complete than those with $\epsilon = 10^{-10}$.

\item It was shown that the scaled mean run-time is inversely 
proportional to $f$ for both accuracy parameters. The basic cause
of this is the specific decrease of the number of bodies.
 
\item Using the model described in section 
\ref{section_improve_the_parameters} the parameters of the collision 
computed from the simulations can be improved. Comparing the improved 
values it turned out that there are not too significant difference 
between the distributions, i.e. from a statistical point of view 
the lower accuracy simulations already provide useful statistical 
data.
\end{itemize}

\section*{Acknowledgements}

This work was partly supported by the ÚNKP-19-4 New National 
Excellence Program of the Ministry for Innovation and Technology, 
partly by the János Bolyai Research Scholarship of the Hungarian 
Academy of Sciences. I acknowledge the support of the Hungarian 
OTKA Grant No. 119993 and the joint OeAD-OMAA program through 
project 95öu10. I would like to thank the support of NVIDIA 
Corporation with the donation of a Tesla C2075 and K40 GPUs. 
The $N$-body calculations were run using the NIIF supercomputer 
facility.








\appendix

\section{Simulation summary}

\begin{table*}
\centering
\caption{Summary of simulation outcomes and results. Each row corresponds 
to one single run data, while the run with 1 - 10 to multiple run data 
and gives the range within which the sample values fall for 
$\epsilon = 10^{-10}$ and $\epsilon = 10^{-13}$, $n$ is the number of collision.}
\label{Tab_A1}
\begin{tabular}{rrrrrrrrrrrr}
\hline  
 & & \multicolumn{5}{c}{$\epsilon = 10^{-10}$} & \multicolumn{5}{c}{$\epsilon = 10^{-13}$} \\ 
\hline  
$f$& run& $n$ & min($V_{\mathrm{i}}'$) & max($V_{\mathrm{i}}'$) & min($Q_{\mathrm{R}}'$) & max($Q_{\mathrm{R}}'$) & $n$  & min($V_{\mathrm{i}}'$) & max($V_{\mathrm{i}}'$) & min($Q_{\mathrm{R}}'$) & max($Q_{\mathrm{R}}'$) \\
                  &    &     & [$V_{\mathrm{esc}}$]  & [$V_{\mathrm{esc}}$]  &                [J kg$^{-1}$] &                [J kg$^{-1}$] &      & [$V_{\mathrm{esc}}$]  & [$V_{\mathrm{esc}}$]  &                [J kg$^{-1}$] &          [J kg$^{-1}$] \\
\hline
\hline
 1  &  1 &  9991 & 0.993 &10.357 & 1.25$\times 10^{4}$ & 4.34$\times 10^{6}$ &  9988 & 0.993 & 8.661 & 1.11$\times 10^{4}$ & 3.58$\times 10^{6}$ \\
 1  &  2 &  9993 & 0.995 & 9.803 & 1.24$\times 10^{4}$ & 5.02$\times 10^{6}$ &  9988 & 0.991 & 7.950 & 1.25$\times 10^{4}$ & 3.72$\times 10^{6}$ \\
 1  &  3 &  9993 & 0.993 & 8.313 & 1.52$\times 10^{4}$ & 5.13$\times 10^{6}$ &  9991 & 0.991 & 7.552 & 1.35$\times 10^{4}$ & 4.18$\times 10^{6}$ \\
 1  &  4 &  9990 & 0.978 & 8.624 & 1.28$\times 10^{4}$ & 4.50$\times 10^{6}$ &  9992 & 0.978 & 8.417 & 1.24$\times 10^{4}$ & 5.69$\times 10^{6}$ \\
 1  &  5 &  9991 & 0.993 & 8.184 & 1.23$\times 10^{4}$ & 4.05$\times 10^{6}$ &  9990 & 0.991 & 9.030 & 1.09$\times 10^{4}$ & 3.52$\times 10^{6}$ \\
 1  &  6 &  9991 & 0.994 &19.517 & 1.64$\times 10^{4}$ & 1.31$\times 10^{7}$ &  9991 & 0.993 & 8.018 & 1.48$\times 10^{4}$ & 5.23$\times 10^{6}$ \\
 1  &  6 &  9990 &       & 8.299 &           & 4.11$\times 10^{6}$ &       &       &       &           &           \\
 1  &  7 &  9988 & 0.988 & 8.222 & 1.23$\times 10^{4}$ & 5.99$\times 10^{6}$ &  9990 & 0.991 & 8.374 & 1.66$\times 10^{4}$ & 3.46$\times 10^{6}$ \\
 1  &  8 &  9990 & 0.992 & 8.017 & 1.22$\times 10^{4}$ & 3.67$\times 10^{6}$ &  9990 & 0.990 & 7.264 & 1.26$\times 10^{4}$ & 4.83$\times 10^{6}$ \\
 1  &  9 &  9989 & 0.985 & 9.190 & 1.18$\times 10^{4}$ & 3.82$\times 10^{6}$ &  9989 & 0.992 & 8.609 & 1.19$\times 10^{4}$ & 3.93$\times 10^{6}$ \\
 1  & 10 &  9990 & 0.995 & 8.111 & 1.33$\times 10^{4}$ & 3.51$\times 10^{6}$ &  9990 & 0.992 & 9.346 & 1.17$\times 10^{4}$ & 6.83$\times 10^{6}$ \\
 1  & 1 - 10 & 99906 & 0.978 &19.517 & 1.18$\times 10^{4}$ & 1.31$\times 10^{7}$ & 99899 & 0.978 & 9.346 & 1.09$\times 10^{4}$ & 6.83$\times 10^{6}$ \\
 1  & 1 - 10 & 99905 &       &10.357 &  & 5.99$\times 10^{6}$ &       &       &       &           &           \\
\hline
 2  &  1 &  9989 & 0.700 & 7.627 & 7.75$\times 10^{3}$ & 2.07$\times 10^{6}$ &  9991 & 0.693 & 7.338 & 7.44$\times 10^{3}$ & 4.33$\times 10^{6}$ \\
 2  &  2 &  9990 & 0.697 & 8.921 & 7.96$\times 10^{3}$ & 3.28$\times 10^{6}$ &  9989 & 0.697 & 9.138 & 6.44$\times 10^{3}$ & 2.86$\times 10^{6}$ \\
 2  &  3 &  9991 & 0.677 & 9.099 & 8.90$\times 10^{3}$ & 2.84$\times 10^{6}$ &  9987 & 0.696 & 7.492 & 7.43$\times 10^{3}$ & 2.03$\times 10^{6}$ \\
 2  &  4 &  9990 & 0.691 & 7.795 & 8.37$\times 10^{3}$ & 2.38$\times 10^{6}$ &  9989 & 0.674 & 9.653 & 6.71$\times 10^{3}$ & 3.20$\times 10^{6}$ \\
 2  &  5 &  9991 & 0.693 & 7.953 & 6.28$\times 10^{3}$ & 2.62$\times 10^{6}$ &  9991 & 0.693 & 8.710 & 7.64$\times 10^{3}$ & 2.60$\times 10^{6}$ \\
 2  &  6 &  9992 & 0.696 & 7.412 & 7.61$\times 10^{3}$ & 2.01$\times 10^{6}$ &  9985 & 0.695 & 7.502 & 7.06$\times 10^{3}$ & 2.02$\times 10^{6}$ \\
 2  &  7 &  9988 & 0.696 & 8.028 & 7.53$\times 10^{3}$ & 2.63$\times 10^{6}$ &  9992 & 0.692 & 8.411 & 6.20$\times 10^{3}$ & 2.43$\times 10^{6}$ \\
 2  &  8 &  9992 & 0.696 & 8.410 & 7.19$\times 10^{3}$ & 2.43$\times 10^{6}$ &  9989 & 0.696 & 8.519 & 7.66$\times 10^{3}$ & 3.95$\times 10^{6}$ \\
 2  &  9 &  9991 & 0.648 & 8.856 & 7.23$\times 10^{3}$ & 2.69$\times 10^{6}$ &  9989 & 0.632 & 8.341 & 7.42$\times 10^{3}$ & 2.39$\times 10^{6}$ \\
 2  & 10 &  9992 & 0.694 & 8.577 & 7.88$\times 10^{3}$ & 3.70$\times 10^{6}$ &  9988 & 0.692 & 7.500 & 7.27$\times 10^{3}$ & 3.76$\times 10^{6}$ \\
 2  & 1 - 10 & 99906 & 0.648 & 9.099 & 6.28$\times 10^{3}$ & 3.70$\times 10^{6}$ & 99890 & 0.632 & 9.653 & 6.20$\times 10^{3}$ & 4.33$\times 10^{6}$ \\
\hline
 3  &  1 &  9993 & 0.565 & 7.856 & 5.61$\times 10^{3}$ & 2.12$\times 10^{6}$ &  9990 & 0.561 & 7.791 & 4.21$\times 10^{3}$ & 2.08$\times 10^{6}$ \\
 3  &  2 &  9990 & 0.565 & 8.372 & 5.43$\times 10^{3}$ & 2.40$\times 10^{6}$ &  9991 & 0.559 & 7.646 & 5.15$\times 10^{3}$ & 2.01$\times 10^{6}$ \\
 3  &  3 &  9991 & 0.560 & 7.478 & 4.69$\times 10^{3}$ & 3.00$\times 10^{6}$ &  9989 & 0.559 & 8.105 & 5.20$\times 10^{3}$ & 2.25$\times 10^{6}$ \\
 3  &  4 &  9989 & 0.560 & 7.793 & 6.08$\times 10^{3}$ & 2.08$\times 10^{6}$ &  9987 & 0.541 & 8.416 & 5.39$\times 10^{3}$ & 2.43$\times 10^{6}$ \\
 3  &  5 &  9990 & 0.564 & 8.490 & 4.60$\times 10^{3}$ & 2.47$\times 10^{6}$ &  9987 & 0.563 & 7.462 & 4.76$\times 10^{3}$ & 3.82$\times 10^{6}$ \\
 3  &  6 &  9988 & 0.565 & 8.362 & 5.09$\times 10^{3}$ & 2.40$\times 10^{6}$ &  9990 & 0.563 & 8.362 & 4.54$\times 10^{3}$ & 2.40$\times 10^{6}$ \\
 3  &  7 &  9989 & 0.563 & 8.900 & 6.00$\times 10^{3}$ & 2.72$\times 10^{6}$ &  9988 & 0.563 & 9.958 & 5.49$\times 10^{3}$ & 3.40$\times 10^{6}$ \\
 3  &  8 &  9991 & 0.561 & 7.210 & 6.26$\times 10^{3}$ & 1.78$\times 10^{6}$ &  9991 & 0.558 & 7.403 & 6.40$\times 10^{3}$ & 2.08$\times 10^{6}$ \\
 3  &  9 &  9989 & 0.540 & 8.358 & 4.40$\times 10^{3}$ & 2.40$\times 10^{6}$ &  9988 & 0.514 & 8.185 & 4.83$\times 10^{3}$ & 2.30$\times 10^{6}$ \\
 3  & 10 &  9990 & 0.548 & 8.371 & 6.59$\times 10^{3}$ & 2.40$\times 10^{6}$ &  9991 & 0.557 & 7.290 & 5.25$\times 10^{3}$ & 1.82$\times 10^{6}$ \\
 3  & 1 - 10 & 99900 & 0.540 & 8.900 & 4.40$\times 10^{3}$ & 3.00$\times 10^{6}$ & 99892 & 0.514 & 9.958 & 4.21$\times 10^{3}$ & 3.82$\times 10^{6}$ \\
\hline
 5  &  1 &  9989 & 0.426 & 7.915 & 2.80$\times 10^{3}$ & 2.15$\times 10^{6}$ &  9986 & 0.427 & 8.430 & 2.72$\times 10^{3}$ & 2.44$\times 10^{6}$ \\
 5  &  2 &  9990 & 0.422 & 8.549 & 3.08$\times 10^{3}$ & 2.51$\times 10^{6}$ &  9987 & 0.429 & 9.260 & 2.68$\times 10^{3}$ & 2.94$\times 10^{6}$ \\
 5  &  3 &  9988 & 0.428 & 7.527 & 3.16$\times 10^{3}$ & 1.94$\times 10^{6}$ &  9985 & 0.429 & 9.100 & 2.76$\times 10^{3}$ & 2.84$\times 10^{6}$ \\
 5  &  4 &  9990 & 0.391 & 8.837 & 3.48$\times 10^{3}$ & 2.68$\times 10^{6}$ &  9991 & 0.397 & 7.775 & 3.60$\times 10^{3}$ & 2.07$\times 10^{6}$ \\
 5  &  5 &  9991 & 0.429 & 8.513 & 3.73$\times 10^{3}$ & 2.49$\times 10^{6}$ &  9989 & 0.424 & 7.451 & 3.14$\times 10^{3}$ & 1.90$\times 10^{6}$ \\
 5  &  6 &  9988 & 0.423 & 8.992 & 3.96$\times 10^{3}$ & 2.77$\times 10^{6}$ &  9989 & 0.423 & 8.356 & 2.42$\times 10^{3}$ & 2.40$\times 10^{6}$ \\
 5  &  7 &  9989 & 0.384 & 7.643 & 3.51$\times 10^{3}$ & 2.00$\times 10^{6}$ &  9990 & 0.422 & 7.637 & 3.69$\times 10^{3}$ & 2.00$\times 10^{6}$ \\
 5  &  8 &  9990 & 0.425 & 7.762 & 3.82$\times 10^{3}$ & 2.22$\times 10^{6}$ &  9991 & 0.424 & 7.757 & 3.63$\times 10^{3}$ & 2.06$\times 10^{6}$ \\
 5  &  9 &  9989 & 0.429 & 9.954 & 3.62$\times 10^{3}$ & 3.40$\times 10^{6}$ &  9988 & 0.425 & 8.065 & 2.70$\times 10^{3}$ & 2.23$\times 10^{6}$ \\
 5  & 10 &  9990 & 0.425 & 8.322 & 3.72$\times 10^{3}$ & 2.38$\times 10^{6}$ &  9989 & 0.428 & 7.507 & 2.96$\times 10^{3}$ & 1.93$\times 10^{6}$ \\
 5  & 1 - 10 & 99894 & 0.384 & 9.954 & 2.80$\times 10^{3}$ & 3.40$\times 10^{6}$ & 99885 & 0.397 & 9.260 & 2.42$\times 10^{3}$ & 2.94$\times 10^{6}$ \\
\hline
10  &  1 &  9990 & 0.287 & 9.232 & 1.54$\times 10^{3}$ & 2.92$\times 10^{6}$ &  9988 & 0.284 & 8.362 & 2.29$\times 10^{3}$ & 2.40$\times 10^{6}$ \\
10  &  2 &  9990 & 0.288 & 8.667 & 2.06$\times 10^{3}$ & 2.58$\times 10^{6}$ &  9989 & 0.287 & 8.667 & 1.73$\times 10^{3}$ & 2.58$\times 10^{6}$ \\
10  &  3 &  9990 & 0.284 & 7.525 & 1.98$\times 10^{3}$ & 1.94$\times 10^{6}$ &  9988 & 0.272 & 7.520 & 2.25$\times 10^{3}$ & 1.94$\times 10^{6}$ \\
10  &  4 &  9990 & 0.247 & 7.763 & 2.09$\times 10^{3}$ & 2.07$\times 10^{6}$ &  9989 & 0.238 & 7.804 & 1.33$\times 10^{3}$ & 2.09$\times 10^{6}$ \\
10  &  5 &  9990 & 0.289 & 8.246 & 1.96$\times 10^{3}$ & 2.33$\times 10^{6}$ &  9991 & 0.286 & 8.373 & 1.49$\times 10^{3}$ & 2.40$\times 10^{6}$ \\
10  &  6 &  9992 & 0.288 & 8.358 & 1.60$\times 10^{3}$ & 2.40$\times 10^{6}$ &  9990 & 0.282 & 8.355 & 1.44$\times 10^{3}$ & 2.39$\times 10^{6}$ \\
10  &  7 &  9990 & 0.291 & 8.580 & 1.59$\times 10^{3}$ & 2.53$\times 10^{6}$ &  9989 & 0.282 & 8.576 & 1.87$\times 10^{3}$ & 2.52$\times 10^{6}$ \\
10  &  8 &  9990 & 0.288 & 8.020 & 1.30$\times 10^{3}$ & 2.21$\times 10^{6}$ &  9989 & 0.286 & 7.004 & 1.91$\times 10^{3}$ & 1.68$\times 10^{6}$ \\
10  &  9 &  9989 & 0.286 & 8.064 & 1.71$\times 10^{3}$ & 2.23$\times 10^{6}$ &  9988 & 0.290 & 8.065 & 1.54$\times 10^{3}$ & 2.23$\times 10^{6}$ \\
10  & 10 &  9990 & 0.200 & 8.326 & 2.05$\times 10^{3}$ & 2.38$\times 10^{6}$ &  9988 & 0.283 & 8.325 & 1.90$\times 10^{3}$ & 2.38$\times 10^{6}$ \\
10  & 1 - 10 & 99901 & 0.200 & 9.232 & 1.30$\times 10^{3}$ & 2.92$\times 10^{6}$ & 99889 & 0.238 & 8.667 & 1.33$\times 10^{3}$ & 2.58$\times 10^{6}$ \\
\hline
\end{tabular}
\end{table*}


\bsp	
\label{lastpage}
\end{document}